\documentclass[12pt]{article}
\usepackage{amsmath}
\usepackage{mathtools}
\usepackage{amssymb}
\usepackage{physics}
\usepackage{ytableau}
\usepackage{braket}
\usepackage{float}
\numberwithin{equation}{section}
\usepackage{cite}
\usepackage{comment}
\pdfoutput=1
\usepackage[pdftex]{graphicx}

\begin{document}
\setlength{\baselineskip}{0.7cm}
\begin{titlepage}
%%%%% PREPRINT NUMBERS %%%%%%
\begin{flushright}
%OMU-PHYS 603  \\
NITEP 227 \\
\end{flushright}
\vspace*{10mm}%
%%%%%%%%%%%%%%%%%%% TITLE %%%%%%%%%%%%%%%%%%
\begin{center}{\Large\bf
Electroweak Symmetry Breaking \\
in Sp(6) Gauge-Higgs Unification Model %on $S^1/Z_2$ \\
%\vspace*{2mm}
%as a Two Higgs Doublet Model 
}
\end{center}
%%%%%%%%%%%%%%%% AUTHORS %%%%%%%%%%%%%%%%%%%%%%%
\vspace*{10mm}
\begin{center}
{\large Nobuhito Maru}$^{a,b}$ and
{\large Akio Nago}$^{a}$ 
\end{center}
%%%%%%%%%%%%%%%%%%%%%%% AFFILIATION %%%%%%%%%%%%
\vspace*{0.2cm}
\begin{center}
%\small
${}^{a}${\it
Department of Physics, Osaka Metropolitan University, \\
Osaka 558-8585, Japan}
\\
%\\[0.2cm]
${}^{b}${\it Nambu Yoichiro Institute of Theoretical and Experimental Physics (NITEP), \\
Osaka Metropolitan University,
Osaka 558-8585, Japan}
%\\[0.2cm}
%${}^{c}${\it}
%%%%%
\end{center}
%%%%%%%%%%%%%%%%%% ABSTRACT %%%%%%%%%%%%%%%
\vspace*{1cm}

\begin{abstract}
We study the electroweak symmetry breaking in a five dimensional $Sp(6)$ gauge-Higgs unification model 
where the weak mixing angle is predicted to be $\sin^2 \theta_W=1/4$ at the compactification scale. 
We find that the correct pattern of electroweak symmetry breaking %and a viable Higgs mass are 
is realized by introducing a 4-rank totally symmetric representation additionally. 
Furthermore, one-loop renormalization group equation analysis show 
that the weak mixing angle at the weak scale 
%is almost the same as that at the compactification scale 
can be in good agreement with the experimental data 
by introducing some fermions in the adjoint representation. 
\end{abstract}
  
\end{titlepage}

\newpage
%%%%%%%%%%%%%%%%%%%%%%%%%%%%%%%%%%%%%%%%%%
\section{Introduction}
%%%%%%%%%%%%%%%%%%%%%%%%%%%%%%%%%%%%%%%%%%
Gauge-Higgs unification (GHU) \cite{Fairlie, Manton, Hosotani} is one of the solutions to the gauge hierarchy problem, 
where the Standard Model (SM) Higgs field is embedded into 
the spatial component of the higher dimensional gauge field. 
A remarkable features of this scenario is that the quantum corrections to the Higgs mass and Higgs potential 
become independent of the cutoff scale of the theory and finite due to the non-local nature of the potential 
\cite{HIL, SSS, ABQ, MY, HMTY, LMH, HLM}. 
In order to embed the SM Higgs field into the gauge field in higher dimensions, 
the SM gauge group is necessary to be extended to some simple group. 
In this extension, since the $U(1)$ hypercharge in the SM is embedded into a simple group, 
the weak mixing angle is also predicted in general. 
In a simplest case of $SU(3)$ gauge theory, 
the SM Higgs field is embedded into the adjoint representation of $SU(3)$, 
which predicts a wrong weak mixing angle $\sin^2 \theta_W=3/4$ \cite{Fairlie, Manton}. 
We notice that there are two 2-rank tensor representations other than the adjoint representation in $SU(3)$, 
namely, 2-rank (anti-)symmetric representations. 
If we consider the cases that the SM Higgs field is embedded into the 2-rank (anti-)symmetric representation 
in $SU(3)$ as a subgroup of $Sp(6)$ \cite{HLM} ($G_2$ \cite{Fairlie, Manton, CGM}) gauge group, 
the weak mixing angle has been known to be predicted such as $\sin^2\theta_W = 1/4$. 

In this paper, we study the electroweak symmetry breaking %and Higgs mass 
in a five dimensional (5D) $Sp(6)$ Gauge-Higgs Unification \cite{HLM}. 
In GHU, Higgs potential at tree level is forbidden by the 5D gauge symmetry 
and the non-local potential is generated by quantum corrections as mentioned above.   
In general, we cannot realize a realistic electroweak symmetry breaking 
unless we introduce additional 5D fermions which do not include SM fermions. 
After studying the properties of the Higgs potential in general, 
we will show that the correct pattern of the electroweak symmetry breaking can be realized 
by introducing a 4-rank totally symmetric representation of $Sp(6)$.   
Furthermore, 
one-loop renormalization group equation analysis show 
that the weak mixing angle at the weak scale 
%is almost the same as that at the compactification scale 
can be in good agreement with the experimental data 
by introducing some fermions in the adjoint representation. 
%it will be shown that the weak mixing angle at the weak scale 
%is almost the same as that at the compactification scale 
%is remarkably good agreement with the experimental data 
%by the 1-loop renormalization group equation analysis. 
%Furthermore, it will be shown that the weak mixing angle at the weak scale is remarkably good agreement 
%with the experimental data by the 1-loop renormalization group equation analysis. 
%In order to obtain 125GeV Higgs boson mass, 
%it will be further shown that the 5D fermions in the adjoint representation of $Sp(6)$ have to be introduced. 

This paper is organized as follows. 
In section 2, we explain our setup. 
In particular, we discuss how the boundary conditions in a extra space are imposed 
and how the SM gauge fields and Higgs field are obtained in our $Sp(6)$ model. 
The general form of one-loop effective potential for Higgs field is calculated in section 3. 
The conditions to realize the correct pattern of the electroweak symmetry breaking are discussed. 
In section 4, based on the argument in the previous section, 
the electroweak symmetry breaking in our model is investigated by introducing additional 5D fermions. 
%Higgs mass analysis is performed in section 5. 
A calculation of the weak mixing angle at the $Z$-boson mass scale 
by the one-loop renormalization group equation is performed in section 5. 
Section 6 is devoted to conclusion.  
We summarize a group theoretical details of $Sp(6)$ in appendices,  
the generators of $Sp(6)$ in appendix A and the decomposition rules of 4-rank totally symmetric representation 
of $Sp(6)$ in appendix B. 

%\newpage
%%%%%%%%%%%%%%%%%%%%%%%%%%%%%%%%%%%%%%%%%%
\section{Setup}
%%%%%%%%%%%%%%%%%%%%%%%%%%%%%%%%%%%%%%%%%%
We consider a 5D $Sp(6)$ GHU model compactified on an orbifold $S^1/Z_2$. 
In general, the weak mixing angle $\theta_W$ can be predicted 
if the hypercharge gauge group $U(1)_Y$ in SM is embedded into a simple group.
In GHU, the minimal simple group is an $SU(3)$, 
where the SM Higgs doublet belongs to the adjoint representation of the $SU(3)$ gauge group. 
However, this model is known to predict a wrong mixing angle $\sin^2 \theta_W = 3/4 $ at the compactification scale. 
In \cite{HLM}, it was discussed that the realistic weak mixing angle $\sin^2 \theta_W = 1/4 \ 
(\sin^2 \theta_W \simeq 0.23$ for an experimental data) can be predicted 
if the SM Higgs doublet belongs to the $SU(3)$ triplet or sextet representation.
A simplest unitary group containing triplet representations under the $SU(3)$ subgroup is an $Sp(6)$, 
 where the adjoint representation of $Sp(6)$ is decomposed in terms of $SU(3)$ representations as %follows, 
 $\bf{21} \rightarrow \bf{8} + \bf{6} + \overline{\bf{6}} + \bf{1}$. 
This is why we consider the $Sp(6)$ gauge group.
A five dimensional spacetime is $M^4 \times S^1/Z_2$, 
where $M^4$ is a four dimensional (4D) Minkowski spacetime 
and $S^1/Z_2$ is a circle with a radius $R$ and divided by a discrete symmetry $Z_2$.
The extra dimensional coordinate of $S^1$ is represented by $x^5 \in [0, 2\pi R)$. 
From
% the periodicity of  $S^1 : x^5 \rightarrow x^5 + 2\pi m R\ (m \in \mathbb{Z})$ and  
the identification by the $Z_2$ transformation $(x^5 \rightarrow -x^5)$, 
the fundamental region of the extra dimension is $x^5 \in [0, \pi R]$ 
and there are two fixed points $x^5=0,\; \pi R$ which are invariant under the $Z_2$ transformation.
Using the $S^1/Z_2$ orbifold, we can consider the parity around the fixed points, which is called $Z_2$ parity.
Furthermore, by choosing the $Z_2$ parity around the fixed point appropriately, 
%well, 
we can construct a chiral theory in a four dimensional theory and realize the gauge symmetry breaking explicitly.

The five dimensional Lagrangian we consider in this paper is given by 
\begin{align}\label{L5}
\mathcal {L}_5 = - \frac{1}{2} \Tr [F^{MN}F_{MN}] - i \bar{\Psi} \Gamma^M D_M \Psi \quad (M,N = 0, 1, 2, 3, 5) ,
\end{align}
where the metric convention is $\eta^{MN} = \text{diag}(+1,-1, \cdots , -1)$ and we denote $\Psi$ as the 5D Dirac fermion.
$\Gamma^M$ are the 5D gamma matrices defined by
\begin{align}
\Gamma^\mu = \left( \mqty{0 & \gamma^\mu \\ \gamma^\mu &0 } \right), \quad
\Gamma^5 = \left( \mqty{0 & i \gamma^5 \\ i\gamma^5 &0 } \right),
\end{align}
where $\mu=0,1,2,3$ is a 4D Lorentz indices. 
The field strength tensor is given by
\begin{align}
&F^{MN} = F^a_{MN}T^a, \notag \\
&F^a_{MN} = \partial_M A^a_N - \partial_N A^a_M - ig_5 {\comm{A_M}{A_N}}^a,
\end{align}
where $T^a = \frac{t^a}{2\sqrt{2}}~(a = 1, \dots , 21)$ is the generators of $Sp(6)$, 
which is summarized in Appendix A. 
%which is the normalized $t^a$ in Appendix A, and the index $a = 1, \dots , 21$ denote the degrees of freedom of Sp(6) gauge. 
$g_5$ is a 5D gauge coupling constant with mass dimension $-1/2$ 
and related to the 4D gauge coupling constant $g$ as $g_5 =\sqrt{\pi R} g$. 
%Note that since the 5D five dimensional Lagrangian is invariant under the $Z_2$ transformation, 
%the boundary conditions in the fixed points of $A_{\mu}$ and $A_5$ differ only in sign.
The boundary conditions of 5D gauge fields at the fixed points are taken as
%that $A_{5}$ is fixed by gauge covariance, 
	\begin{align}
	\label{Z2BC1}
		A_\mu(x^\mu, x^5_i-x^5)=+P_i A_\mu(x^\mu, x^5)P_i^{\dag},\\
		A_5(x^\mu,  x^5_i-x^5)=-P_i A_5(x^\mu, x^5)P_i^{\dag},
		\label{Z2BC2}
	\end{align}
where $P_i \ (i =0, 1)$ is the parity matrix at each fixed point $x^5_0 = 0,\; x^5_1 = \pi R$. 
The boundary condition of $A_\mu$ is chosen by hand so that the electroweak gauge symmetry is unbroken. 
On the other hand, that of $A_5$ is accordingly determined by the covariance of $F_{\mu5}$ under $Z_2$.  
%Here, 
In order to achieve gauge symmetry breaking $Sp(6) \rightarrow SU(2)_L \times U(1)_Y \times U(1)_X$ by $Z_2$ parity of $A_\mu$, 
the parity matrix $P_{0,1}$ is chosen as follows,
	\begin{align}
		&P_0 = \text{diag}( +1, +1, -1, +1, +1, -1),	 \\			
		&P_1 = \text{diag}( +1, +1, +1, -1, -1, -1).
	\end{align}
The combination of the parity matrices is denoted as
	\begin{equation}
		P=\text{diag}\big((++),(++),(-+),(+-),(+-),(--) \big) ,
	\end{equation}
where the left and right $\pm$ of each component of $P$ correspond to the $\pm 1$ of each component of $P_{0,1}$, respectively. 
Note that the product $P_0 P_1 = P_1P_0 = 1(-1)$ corresponds to the (anti-)periodic boundary condition around $S^1$, respectively. 

%This expression gives the $Z_2$ parity of each component of the gauge fields from \eqref{Z2BC1} and \eqref{Z2BC2},
$Z_2$ parity of each component of the gauge fields  can be read from \eqref{Z2BC1} and \eqref{Z2BC2},
\begin{align}\label{gp}
		A_\mu &\rightarrow 
			\left(
			\begin{array}{ccc|ccc}
				(++)&(++)&(-+)&(+-)&(+-)&(--)\\
				(++)&(++)&(-+)&(+-)&(+-)&(--)\\
				(-+)&(-+)&(++)&(--)&(--)&(+-)\\ \hline
				(+-)&(+-)&(--)&(++)&(++)&(-+)\\
				(+-)&(+-)&(--)&(++)&(++)&(-+)\\
				(--)&(--)&(+-)&(-+)&(-+)&(++)\\
			\end{array}\right), \\[7pt]
			\label{hgp}
		A_{5} &\rightarrow 
			\left(
			\begin{array}{ccc|ccc}
				(--)&(--)&(+-)&(-+)&(-+)&(++)\\
				(--)&(--)&(+-)&(-+)&(-+)&(++)\\
				(+-)&(+-)&(--)&(++)&(++)&(-+)\\ \hline
				(-+)&(-+)&(++)&(--)&(--)&(+-)\\
				(-+)&(-+)&(++)&(--)&(--)&(+-)\\
				(++)&(++)&(-+)&(+-)&(+-)&(--)\\
			\end{array}\right).
	\end{align}
%From this $Z_2$ parity we consider the (anti-)periodic boundary condition of $S^1/Z_2$.
The gauge field can be expanded in terms of the Kaluza-Klein (KK) mode functions,
\begin{align}
		A^{++}_M(x^\mu, x^5) &=
		\frac{1}{\sqrt{{\pi R}}} A^{(0)}_M( x^{\mu} ) 
		\ +\  \sqrt{\frac{2}{\pi R}}\ {\displaystyle \sum_{n = 1}^\infty}\  
		 A^{(n)}_M( x^{\mu} )\cos	\left( \frac{n x^5}{R} \right),		\\[5pt]
		A^{--}_M(x^\mu, x^5) &= \hspace*{34mm}
		 \sqrt{\frac{2}{\pi R}}\ {\displaystyle \sum_{n = 1}^\infty}\  
		 A^{(n)}_M( x^{\mu} )\sin	\left( \frac{n x^5}{R} \right),		\\[5pt] 
		A^{+-}_M(x^\mu, x^5) &=\hspace*{34mm}
		 \sqrt{\frac{2}{\pi R}}\ {\displaystyle \sum_{n = 0}^\infty} \ 
		 A^{(n)}_M( x^{\mu} )\cos	\left( \frac{(n+\frac{1}{2}) x^5}{R} \right),	\\[5pt] 
		A^{-+}_M(x^\mu, x^5) &=\hspace*{34mm}
		 \sqrt{\frac{2}{\pi R}}\ {\displaystyle \sum_{n = 0}^\infty} \ 
		 A^{(n)}_M( x^{\mu} )\sin	\left( \frac{(n+\frac{1}{2}) x^5}{R} \right).
	\end{align}
Obviously, $A_M^{(+,+)}(x^\mu, x^5)$ means the 5D gauge field with $Z_2$ parity $(P_0, P_1)=(+,+)$ and so on.  
$A_M^{(n)}(x^\mu)$ is 4D gauge fields of $n$-th KK modes. 
Since the zero mode ($n=0$) corresponding to the 4D effective field appear only in the $(++)$ part, 
the 4D gauge field are $SU(2)_L \times U(1)_Y \times U(1)_X$ gauge fields from \eqref{gp}, 
\begin{align}\label{defgauge}
A_{\mu}^{(0)}= \frac{1}{2}
				\begin{pmatrix}\mqty{
				G_\mu & 0 \\
				0 & G_\mu			
				}
				\end{pmatrix} + X_\mu \frac{1}{2\sqrt{3}}\; \text{diag}(1,1,1,-1,-1,-1),	\\[7mm]%\quad	%\\[4mm]
			G_\mu =
			\begin{pmatrix}\mqty{
			\frac{1}{\sqrt{2}}Z_\mu+\frac{1}{\sqrt{6}}B_\mu&W_\mu^+&0 \\
			W_\mu^-&-\frac{1}{\sqrt{2}}Z_\mu+\frac{1}{\sqrt{6}}B_\mu&0 \\
			0&0&-\frac{2}{\sqrt{6}}B_\mu		
			}
			\end{pmatrix}.
\end{align}
In \eqref{hgp}, a doublet appears in the off-diagonal component of the zero mode of $A_5$, 
which is considered to be the SM Higgs doublet $H$,
\begin{align}\label{defHiggs}
		A_{5}^{(0)}= \frac{1}{2}
				\begin{pmatrix}\mqty{
				0 & \Phi \\
				\Phi^\dagger & 0			
				}
				\end{pmatrix},	\quad	%\\[4mm]
			\Phi=
			\begin{pmatrix}\mqty{
			0&0&\phi^{+}\\
			0&0&\phi^0 \\
			\phi^{+}&\phi^{0}&0			
			}
			\end{pmatrix}, \quad
			H=
			\begin{pmatrix}\mqty{
				\phi^{+}\\
				\phi^0
			}\end{pmatrix}.
	\end{align}
Now we can see that the Higgs doublet is embedded in the sextet of $SU(3)$, which is a subgroup of $Sp(6)$. 
Therefore, the weak mixing angle can be predicted to be $\sin^2 \theta_W = \frac{1}{4}$ at the compactification scale.
Furthermore, $A^{(0)}_5$ can be expanded around the vacuum as follows
\begin{align}
A^{(0)}_5 = \ev{A^{(0)}_5} + {\tilde{A}}^{(0)}_5 .
\end{align}
As in the SM, the neutral Higgs field takes the vacuum expectation value (VEV) :
\begin{align}
			\ev{A^{}_5}=
%			\begin{pmatrix}\mqty{
%			0 \\
%		\frac{\ev{A^{14}_5}}{\sqrt{2}}
%			}\end{pmatrix}
%			=
%			\begin{pmatrix}\mqty{
%				0\\
%				\frac{v}{\sqrt{2}}
%			}\end{pmatrix}.
			\left(
			\begin{array}{ccc|ccc}
				0&0&0&0&0&0\\
				0&0&0&0&0&v\\
				0&0&0&0&v&0\\ \hline
				0&0&0&0&0&0\\
				0&0&v&0&0&0\\
				0&v&0&0&0&0\\
			\end{array}\right) , \quad 
			\ev{H}
			=
			\begin{pmatrix}\mqty{
				0\\
			\frac{\ev{A_5^{14}}}{\sqrt{2}}
			}\end{pmatrix}.
\label{HiggsVEV}
\end{align}
As is seen from Appendix A, the neutral component corresponds to the direction of 14-th $Sp(6)$ generator.  
%Due to the gauge symmetry, the tree-level potential does not appear in this model.
%%%%%%%%%%%%%%%%%%%%%%%%%%%%%%%%%%%%%%%%%%
\section{One-loop effective potential}%Spontaneous symmetry breaking}
%%%%%%%%%%%%%%%%%%%%%%%%%%%%%%%%%%%%%%%%%%
In GHU, the Higgs potential at the tree level is forbidden by %does not appear from 
 the gauge invariance and the VEV of the Higgs field cannot be determined.
In this section, we consider an electroweak symmetry breaking from one-loop effective potential.
%Using background field methods, 
The 4D effective potential is given by
	\begin{align}\label{veff}
		V_{\text{eff}}^{\text{1-loop}} =  \sum_{n}N\frac{(-1)^{F}}{2}
		\int \frac{\dd^4 p}{( 2 \pi )^4} \Tr \log( p^2 + M^2_{n}),
	\end{align}
where the sum is taken over all KK modes with mass $M_{n}  (n \in \mathbb{Z}$). %-\infty < n <\infty)$.
The $F$ denotes the fermion number, $F = 0$ for bosons and $F = 1$ for fermions.
The factor $N$ is the degrees of freedom of the field propagating in the loop.
When the Higgs field takes the VEV, 
the KK mass depends on the Aharanov-Bohm (AB) phase $\theta_{AB}$ produced by the compactification of the extra dimension.
In fact, $\theta_{AB}$ appears as the phase of the eigenvalues of the Wilson line, 
which is a line integral of $\ev{A_5}=\frac{\ev{A^{(0)}_5}}{\sqrt{\pi R}}$ along the $S^1$.
%\begin{align}
%\theta_{AB} = {g}\oint_{S^1} \dd x^5 \frac{\ev{A_5^{14}}}{\sqrt{2}}  ,
%\end{align}
%	\begin{align}
%		V_{\text{eff}}^{\text{1-loop}}
%		&= - \frac{( - 1 )^{F}}{2}N \sum_{n=-\infty}^\infty 
%		\frac{\dd t}{t} \int \frac{d^4 p}{( 2 \pi )^4} e^{- t ( p^2 + M^2_{n_1, n_2}) } \notag \\
%		& = - \frac{( - 1 )^{F}}{32 \pi^2}N \sum_{n=-\infty}^\infty \int_0^\infty \dd{l} 
%		\; le^{-M^2_{n}/l}
%	\end{align}
Now we consider the contribution from a scalar field with a typical mass spectrum
\begin{align}
M_{n}^2 = \left(\frac{ (n+\frac{\beta}{2}) +\frac{m \alpha}{2}}{R}\right)^2 \quad \qty(\alpha = \frac{v g R}{\sqrt{2}},\ m \in \mathbb{Z} ),
%M_{n}^2 = \left( \frac{n + \alpha}{R}\right)^2 \quad (\alpha = v g R),
\end{align}
where the parameter $\beta=0(1)$ comes from the (anti-)periodic boundary condition of the fields, respectively.
% or $\beta=1$ for the anti-periodic one.
$\alpha$ is a dimensionless parameter that characterizes the AB phase and satisfies $2 \pi \alpha=\theta_{AB}$.  
Also, for an integer shift of $\alpha$, the AB phase is periodic. 
This implies that the Wilson line is gauge invariant. 
The factor $\sqrt{2}$ in the denominator of $\alpha$ is due to the normalization constant of the gauge group generator, 
and the integer $m$ is determined by the structure constant of the gauge group.
%We consider the contribution of a field with the Poisson resummation formula
%	\begin{align}
%		\sum_{n=-\infty}^\infty \exp[-\frac{(n+\alpha^{\prime})^2}{he R^2 l}]=R \sqrt{l \pi}\sum_{k=-\infty}^\infty 
%		\exp[2 \pi i k \alpha^{\f\prime} - R^2 l \pi^2 k^2]
%	\end{align}
If we change the sum of the KK mode momentum number $n$ to the winding number $k$, 
we obtain the contribution from the scalar field
	\begin{align}
%		V_{\text{eff}}^{\text{1-loop}} (\alpha)
		\nu^s (\alpha; \beta, m)
		= - \frac{R}{32 \pi^{3/2}} \sum_{k=-\infty}^\infty 
		(-1)^{\beta k} e^{i m \pi  k \alpha} \int_0^\infty \dd{l} \; l^{3/2}  
		e^{- R^2 l \pi^2 k^2 } .
%		V_{\text{eff}}^{\text{1-loop}} 
%		= - \frac{(-1)^F R}{32 \pi^{3/2}}N \sum_{k=-\infty}^\infty 
%		e^{2 \pi i k \alpha} \int_0^\infty \dd{l} \; l^{3/2}  
%		e^{- R^2 l \pi^2 k^2 } .
	\end{align}
The $k=0$ part is %a constant term in the potential 
independent of the VEV of Higgs field parameter $\alpha$, 
which is not considered below because it does not affect the analysis of electroweak symmetry breaking.
%Finally, 
After the change of variable $l \rightarrow l^{\prime} = R^2 l \pi^2 k^2 $ and integration with respect to $l^\prime$, 
the contribution to the one-loop effective potential by a scalar field with the typical KK mass is obtained.
	\begin{align} \label{potential_fomula}
%		V_{\text{eff}}^{\text{1-loop}}(\alpha)
		\nu^s (\alpha; \beta, m)
%		&=- \frac{(-1)^F R}{32 \pi^{3/2}}N \sum_{k=-\infty}^\infty 
%		e^{2 \pi i k \alpha^{\prime}} \int_0^\infty \frac{\dd{l^\prime}}{(R \pi k)^2} \; \frac{{l^\prime}^{3/2}}{(R \pi |k|)^3}  
%		e^{- l^\prime } \notag
%		&= - \frac{(-1)^F \Gamma(\frac{5}{2})}{32 \pi^6 \sqrt{\pi} R^4}N \sum_{ k \neq 0} \frac{(-1)^{\beta k} e^{i m \pi  k \alpha}}{|k|^5} \notag \\
		&= - \frac{3}{64 \pi^6 R^4} \sum_{ k = 1}^\infty \frac{(-1)^{\beta k} \cos( m \pi k \alpha)}{k^5}. 
%		&\equiv - \frac{3(-1)^F}{64 \pi^6 R^4} N \sum_{k = 1}^\infty \nu(\alpha^{\prime}).
%		&= - \frac{(-1)^F \Gamma(\frac{5}{2})}{32 \pi^6 \sqrt{\pi} R^4}N \sum_{ k \neq 0} \frac{ e^{2 \pi i k \alpha}}{|k|^5} \notag \\
%		&= - \frac{3(-1)^F}{64 \pi^6 R^4}N \sum_{ k = 1}^\infty \frac{\cos( 2 \pi k \alpha)}{k^5}. 
	\end{align}
In $Sp(6)$ GHU model, %with a simple compactification of the extra dimension, 
the general effective potential including contribution of various fields can be written as follows 
\begin{align}
V_{\text{eff}}^{\text{1-loop}}(\alpha)
%		&\sim \sum_l \sum_{ k = 1}^\infty  \{ a_{l} + b_{l}(-1)^k \} \mathfrak{Re} \left[F\left(\frac{l \alpha^\prime}{2}, 5 \right) \right]　\notag \\
		&= - \frac{3}{64 \pi^6 R^4} \sum_m \sum_{ k = 1}^\infty  \{ a_{m} + b_{m}(-1)^k \}  \frac{\cos( m \pi k \alpha)}{k^5},
		\label{GeneralPotential}
\end{align}
where $a_m, \ b_m$ are the coefficients associated with the representation of $SU(2)_L$ which is a subgroup of $Sp(6)$ 
and the degeneracy of KK masses. 
%each field and includes the degenerate number of KK mass.
Since this potential is symmetric under the operation $\alpha \rightarrow -\alpha$.
%Futhermore, if $a_m=b_m$ for odd $m$, 
%\begin{align}
%V_{\text{eff}}^{\text{1-loop}}(\alpha+\frac{1}{2}) = V_{\text{eff}}^{\text{1-loop}}(\alpha - \frac{1}{2})
%\end{align}
%the effective potential is symmetric with respect to $\alpha^\prime = \frac{1}{2}$ :
%\begin{align}
%V_{\text{eff}}^{\text{1-loop}}(\alpha+{1}/{2})=V_{\text{eff}}^{\text{1-loop}}(\alpha-{1}/{2}).
%\end{align}
and the Wilson line is invariant under a integer shift of $\alpha$,   
%let us consider the case where the minimum point is in $0 \leq \alpha_{\text{min}} \leq \frac{1}{2}$.
we have only to consider the case where the potential minimum is located 
in the fundamental region $0 \leq \alpha_{\text{min}} \leq \frac{1}{2}$. 

Next, %we have a look at the gauge symmetry that remains in the vacuum.
we study the gauge symmetry breaking. 
The gauge symmetry is spontaneously broken when the gauge field obtain mass proportional to VEV, 
which is given by the commutator of the VEV of Wilson line and the broken generators.
%Thus, 
In other words, the generators of the unbroken symmetries commute with the VEV of the Wilson line:
\begin{align}
\ev{W_{}}
&=\mathcal{P} \exp [i {g_5} \oint_{S^1}  \dd x^{5} \ev{A_{5}}]
%&=\exp[i \pi \alpha_{1,2}\: t^{11}] 
%\notag \\[5mm]
%&=P \exp [2 \pi i \alpha^\prime t^{14}] 
\notag \\[5mm]
&=\left(\begin{array}{ccc|ccc}
1 & 0 & 0 & 0 &0 &0\\
0 & \cos(2\pi \alpha) &  0 & 0 &0 &i \sin(2\pi \alpha)  \\
0 & 0 &  \cos(2\pi \alpha) & 0 &i \sin(2\pi \alpha)  &0\\ \hline
0 & 0 & 0 & 1 &0 &0\\ 
0 & 0 &i \sin(2\pi \alpha)  & 0 & \cos(2\pi \alpha) & 0 \\
0 &i \sin(2\pi \alpha)  & 0 & 0 & 0 & \cos(2\pi \alpha)
\end{array} 
\right).
\end{align}
The symmetry breaking patterns can be classified into the following three cases. 

\noindent (i) $\alpha_{\text{min}}= 0$
\begin{align}
\ev{W_{}}=I_6
\end{align}
Since $\ev{W}$ is a $6 \times 6$ identity matrix $I_6$, it is commutative with all generators of $Sp(6)$ 
and the electroweak symmetry breaking do not occur.
\begin{align}
Sp(6) \xrightarrow{\text{$Z_2$ orbifold}} SU(2)_L \times U(1)_Y \times U(1)_X.  
\end{align}\\[-12pt]
(ii) $\alpha_{\text{min}}=\frac{1}{2}$
\begin{align}
\ev{W_{}}= \text{diag}(+1, -1, -1, +1, -1, -1)
\end{align}
Since $\ev{W}$ is diagonal, it is commutative with the diagonal generators of $Sp(6)$, namely $T^3,T^8$ and $T^9$:
%\begin{align}
%\ev{W}&=\left(\begin{array}{ccc|ccc}
%1 & 0 & 0 & 0 &0 &0\\
%0 & -1 &  0 & 0 &0 &0  \\
%0 & 0 & -1 & 0 &0  &0\\ \hline
%0 & 0 & 0 & 1 &0 &0\\ 
%0 & 0 & 0  & 0 &-1 & 0 \\
%0 & 0  & 0 & 0 & 0 & -1
%\end{array} 
%\right)
%\end{align}
%である。
\begin{align}
[\ev{W}, T^3]=[\ev{W}, T^8]= [\ev{W}, T^9]= 0.
\end{align}
Thus, only three U(1) symmetries remain unbroken, and the following gauge symmetry breaking takes place, 
\begin{align}
Sp(6) \xrightarrow{\text{$Z_2$ orbifold}}SU(2)_L \times U(1)_Y \times U(1)_X \xrightarrow{\text{quantum effect}}U(1) 
\times U(1)_Y \times U(1)_X. 
\end{align}\\[-12pt]
(iii) $0 < \alpha_{\text{min}} <\frac{1}{2} \\%,\ \frac{1}{2} < \alpha^\prime_{\text{min}} < 1
$
In this case, we find the following commutation relations. 
\begin{align}
&[\ev{W}, T^3]
%= i \frac{\sin(2\pi \alpha^\prime)}{2\sqrt{2}} \left(\begin{array}{ccc|ccc}
%0 & 0 & 0 & 0 &0 &0\\
%0 & 0 &  0 & 0 &0 & 1 \\
%0 & 0 &  0 & 0 & 1 &0\\ \hline
%0 & 0 & 0 & 0 &0 &0\\ 
%0 & 0 & - 1 & 0 & 0 & 0 \\
%0 & - 1& 0 & 0 & 0 &0
%\end{array} 
%\right) 
= -\sin(2\pi \alpha_{\text{min}})T^{21},\quad 
[\ev{W},T^8] = -\frac{\sin(2\pi \alpha_{\text{min}})}{\sqrt{3}}T^{21}, \notag \\
&[\ev{W},T^9] = \frac{2 \sqrt{2}\sin(2\pi \alpha_{\text{min}})}{\sqrt{3}}T^{21} .
\end{align}
Some linear combinations of $T^3,T^8$ and $T^9$ commute with $\ev{W}$.
$\frac{1}{2}T^3-\frac{\sqrt{3}}{2}{T^8}$ is identified with $U(1)_{em}$ generator and commutes with $\ev{W}$.
In this case, the electroweak symmetry breaking similar to the SM can be achieved.
\begin{align}
Sp(6) \xrightarrow{Z_2\: \text{orbifold}} SU(2)_L \times U(1)_Y 
\times U(1)_X \xrightarrow{\text{quantum effect}} U(1)_{em} \times U^\prime(1). 
\end{align}
Here, the generator of $U(1)'$ is defined as 
$\frac{1}{\sqrt{2}} \left( \frac{1}{2}T^3+\frac{\sqrt{3}}{2}{T^8} \right) + \frac{1}{\sqrt{2}}T^9$. 

%%%%%%%%%%%%%%%%%%%%%%%%%%%%%%%%%%%%%%%%%%
\section{Analysis of electroweak symmetry breaking} %in $Sp(6)$ Model}
%%%%%%%%%%%%%%%%%%%%%%%%%%%%%%%%%%%%%%%%%%
%\begin{align}
%V_{\text{eff}}
%		&\sim \sum_l \sum_{ k = 1}^\infty  \{ a_{l} + b_{l}(-1)^k \} \mathfrak{Re} \left[F\left(\frac{l \alpha^\prime}{2}, 5 \right) \right] \notag \\
%		&= \sum_l \sum_{ k = 1}^\infty  \frac{1}{k^5}\left[ \{ a_{2l} + b_{2l}(-1)^k \}\cos( 2 l \pi k \alpha^\prime) + \{a_{2l-1} + b_{2l-1}(-1)^k \} \cos((2l-1)\pi k \alpha^\prime)\right]
%\end{align}
In the GHU, the potential minimum is determined at the quantum level, %since the Higgs field has gauge symmetry.
since the non-local potential is generated by the quantum corrections although the local potential is not allowed by the gauge invariance. 
If the potential minimum $\alpha_{\text{min}}$ is non-zero, the electroweak symmetry is spontaneously broken 
and the curvature of the potential at $\alpha = 0$ is negative:
\begin{align}
		 \eval{\pdv{^2}{{\alpha}^2} V^{\text{1-loop}}_{\text{eff}}}_{\alpha=0} %\sim -\sum_{m} m^2\left(a_{m} -  \frac{3b_{m}}{4}\right) \zeta(3)  
		< 0. 
\end{align}
In this case, the relation between the coefficients of the potential \eqref{GeneralPotential} is given by
\begin{equation}\label{EWSB}
\sum_{m} m^2 a_m < \frac{3}{4}\sum_{m} m^2 b_m .
\end{equation}

%%%%%%%%%%%%%%%%%%%%%%%%%%%%%%%%%%%%%%%%%%%%
\subsection{$Sp(6)$ gauge fields}
\begin{figure}[H]
		\centering
		\includegraphics[width=100mm]{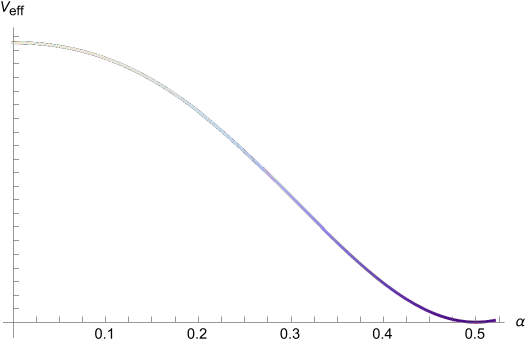}
		\caption{Effective potential of Sp(6) gauge fields $V^{\text{g}}_{\text{eff}}$.
		The potential minimum is found to be $\alpha_{\text{min}}=\frac{1}{2}$.}
		\label{fig:potentialplot0}
	\end{figure}

%%%%%%%%%%%%%%%%%%%%%%%%%%%%%%%%%%%%%%%%%%
To calculate the one-loop effective potential for the gauge field from \eqref{potential_fomula}, we only need to find its KK mass spectrum.
Since the Higgs field appears in the extra-dimensional component of the higher-dimensional gauge field in GHU, 
 the KK mass dependent on the VEV is given by the covariant derivative in the extra-dimensional direction.
The squared mass matrix $\hat{M}^2$ of the gauge field is given by 
\begin{align}
\hat{M}^2=-{D_5^2},\quad  D_5^{ab} = \delta^{ab}\partial_\mu - i g f^{acb}A^c_5 ,
\end{align} 
where $f^{acb}$ is structure constant of $Sp(6)$. 
From \eqref{HiggsVEV}, since the fifth component of the gauge field corresponding to the $Sp(6)$ generator $T^{14}$ takes the VEV, 
we search for the generators that do not commute with $T^{14}$.
%\begin{align}
%\comm{T^a}{T^b}=-if^{abc}T^c
%\end{align}
The non-zero structure constants of type $f^{a\; 14\; b}$ are as follows,    
	\begin{align}
		\begin{cases}
			f^{1\, 14\, 21} = f^{4\, 14\, 19} =f^{13\, 14\, 5} = f^{15\, 14\, 2} = f^{20\, 14\, 3}= \displaystyle\frac{1}{2\sqrt{2}}, 	\\[3mm]
			f^{6\, 14\, 17} = f^{6\, 14\, 18}= f^{7\, 14\, 12}= f^{11\, 14\, 7}= \displaystyle\frac{1}{2}, 	\\[3mm]
			f^{20\, 14\, 8}=\displaystyle\frac{1}{2 \sqrt{6}},\\[3mm]
			  f^{9\, 14\, 20}=\displaystyle\frac{1}{ \sqrt{3}}.
		\end{cases}
	\end{align}
By using the mode expansions and diagonalizing the above mass matrix, 
we can obtain the KK mass $M_n^2$ %that depend on the VEV is
%\begin{align}
%		M^2_n= &\frac{n^2}{R^2} \times 3,\; \frac{(n+\frac{1}{2})^2}{R^2} \times 4,\; \frac{\left( n \pm \frac{\alpha}{\sqrt{2}}\right)^2 }{R^2} \times 1,\; \frac{\left( n \pm \frac{\alpha}{2\sqrt{2}}\right)^2 }{R^2} \times 2,\notag \\[2mm]
%		& \frac{\left\{(n + \frac{1}{2}) \pm \frac{\alpha}{\sqrt{2}} \right\}^2 }{R^2} \times 2,\;\frac{\left\{(n + \frac{1}{2}) \pm \frac{\alpha}{2\sqrt{2}} \right\}^2 }{R^2} \times 2 \qquad ( -\infty <n<\infty) .
%	\end{align}
%Therefore, the KK mass that depend on VEV is,
	\begin{align}
		M^2_n= 
		&\left( \frac{n + \alpha }{R} \right)^2 \times 1,\; 
		\left( \frac{(n + \frac{1}{2}) + \alpha }{R} \right)^2 \times 2, \notag \\[3mm]
		&\left( \frac{n + \frac{\alpha }{2}}{R} \right)^2 \times 2,\; 
		\left( \frac{(n + \frac{1}{2}) + \frac{\alpha }{2}}{R} \right)^2 \times 2 \qquad ( -\infty <n<\infty) .
		\end{align}
Taking into account the degree of freedom of the massive gauge field, %which is 3, 
the one-loop effective potential of the $Sp(6)$ gauge field is obtained from \eqref{potential_fomula},
	\begin{align}
		V_{\text{eff}}^g (\alpha)
		&= \nu^g (\alpha \ ; 0, 2) + 2 \nu^g (\alpha \ ; 1, 2)+ 2 \nu^g (\alpha \ ; 0, 1)+ 2 \nu^g (\alpha \ ; 1, 1) \notag \\
		&=-\frac{9}{64 \pi^6 R^4}  \sum_{ k = 1}^\infty \frac{1}{k^5}\qty[ \{ 1 + 2(-1)^k \}\cos( 2 \pi k \alpha) + 2\{1 + (-1)^k \} \cos( \pi k \alpha)] .
%		&=-\frac{9}{64 \pi^6 R^4}  \sum_{ k = 1}^\infty  \frac{1}{k^5}\left[ \{ 1 + 2(-1)^k \}
%\cos( \sqrt{2} \pi k \alpha) + 2\{1 + (-1)^k \} \cos( \frac{\pi k \alpha}{\sqrt{2}})\right]
	\end{align}
%Futhermore, if $a_m=b_m$ for odd $m$,
%\begin{align}
%		 \eval{\pdv{}{\alpha} V^{\text{1-loop}}_{\text{eff}}}_{\alpha=\frac{1}{2}} %\sim -\sum_{m} m^2\left(a_{m} 
%-  \frac{3b_{m}}{4}\right) \zeta(3)  		= 0 .
%\end{align}
The curvature of the potential at $\alpha = 0$ should be negative:
\begin{align}
		 \eval{\pdv{^2}{{\alpha}^2} V^{g}_{\text{eff}}}_{\alpha=0} %\sim -\sum_{m} m^2\left(a_{m} -  \frac{3b_{m}}{4}\right) \zeta(3) 
		&=\frac{9}{64 \pi^4 R^4}  \sum_{ k = 1}^\infty \frac{1}{k^3} \left\{ 6 + 10(-1)^k \right\} \notag \\[2mm]
%		&=\frac{9}{64 \pi^4 R^4} \left\{ 6\: \zeta(3)+ 10\: Li_3(e^{i \pi}) \right\}\notag \\[2mm]
		&=\frac{9}{64 \pi^4 R^4} \left( 6 -  \frac{15}{2} \right)\zeta(3) < 0.
\end{align}
Thus, the potential has the minimum at $0<\alpha_{\text{min}}\leq 1/2$ and the gauge symmetry is spontaneously broken.
Here, $\zeta(s)$ is the Riemann zeta function
% and $ Li_s(z)$ is polylogarithm:
\begin{align}
\zeta(s)=\sum_{n=1}^\infty \frac{1}{n^s}
%, \quad Li_s(z)= \sum_{n=1}^\infty \frac{z^n}{n^s} 
, \qquad 
%\end{align}
%where $s$%,\ z\ (\abs{z}<1)$ is arbitrary complex number. 
%Furthemore, the two functions are related as follows
%\begin{align}
%Li_s(e^{i\pi}) = 
\sum_{n=1}^\infty \frac{(-1)^n}{n^s}=\qty(-1+\frac{1}{2^{s-1}})\zeta(s),
\end{align}
where $s$ is a complex number with a real part that satisfies $\mathfrak{Re}\: s > 1$.
The effective potential is symmetric with respect to $\alpha=1/2$, because of $a_m=b_m$ for odd $m$.
The curvature of the potential at $\alpha = 1/2$ is found to be positive:
\begin{align}
		 \eval{\pdv{^2}{\alpha^2} V^{\text{1-loop}}_{\text{eff}}}_{\alpha=\frac{1}{2}} %\sim -\sum_{m} m^2\left(a_{m} -  \frac{3b_{m}}{4}\right) \zeta(3)  
		&=\frac{9}{64 \pi^4 R^4}  \sum_{ k = 1}^\infty \frac{1}{k^3}\qty[4\{ 1 + 2(-1)^k \}\cos( \pi k)+ 2\{1 + (-1)^k \} \cos(\frac{ \pi k}{2})] \notag \\[3mm]
		&=\frac{9}{64 \pi^4 R^4}  \qty(5 -\frac{3}{8})\zeta(3) >0
\end{align}
Therefore, the potential minimum is found at $\alpha_{\text{min}}=1/2$ as in Figure \ref{fig:potentialplot0}.
This means that the obtained symmetry breaking is not the electroweak symmetry breaking of our interest from the argument in section 3.
%$\alpha^\prime_{\text{min}}=\frac{1}{2}$, therefore electoroweak symmetry breaking do not occur. 

%%%%%%%%%%%%%%%%%%%%%%%%%%%%%%%%%%%%%%%%%%
\subsection{Fermions}
%%%%%%%%%%%%%%%%%%%%%%%%%%%%%%%%%%%%%%%%%%
%We introduce fermions that can predict realistic Higgs mass 
%while achieving spontaneous electroweak symmetry breaking similar to the Standard Model.
In this subsection, we discuss possibilities that a realistic Higgs mass 
as well as the electroweak symmetry breaking are realized by introducing fermions. 

We would like to comment on the SM fermions in our model before the Higgs potential analysis. 
In the $Sp(6)$ model, the Higgs doublet is embedded in the $SU(3)$ sextet after the decomposition of $Sp(6)$ representation  
in order to predict a realistic weak mixing angle. 
In this case, the electromagnetic charges of the $SU(3)$ triplet is known as $(0, -1, 1)^T~(T:$ Transpose) \cite{HLM} 
where the first two components correspond to $SU(2)_L$ doublet.  
On the other hand, the $SU(2)_L$ singlet is possible to be embedded into the 2-rank anti-symmetric representation of $SU(3)$ in $Sp(6)$.  
Since the electromagnetic charges of $SU(3)$ triplet are integers, any fractional charges cannot be obtained by its tensor product. 
Therefore, the SM quarks are introduced by hand on the 4D brane at the fixed points.  
%The sextet of $SU(3)$ is a 2-rank totally symmetric representation of $SU(3)$.
%The neutral Higgs charge in the sextet of $SU(3)$ must be zero, since the fundamental representation of $SU(3)$ can only contain integer charges.
%Therefore, only the SM lepton can be included in the Sp(6) representation. we assume that the SM quark are localized at a fixed point.
In our setup, Yukawa coupling cannot be obtain from the 5D gauge coupling. 
The SM leptons also have to put on the 4D brane at the fixed points.  
Yukawa coupling can be obtain by introducing additional massive 5D fermions and coupling them 
to the SM fermions through the Dirac mass terms \cite{SSS}. 
In any case, the SM fermions and additional fermions to realize Yukawa couplings are massive. 
Note that the contribution of the one-loop effective potential of the massive fermion is strongly suppressed 
by the Boltzman-like factor $e^{-\pi M R}$, where $M$ is the 5D fermion mass.
%In this paper, we consider the adjoint representation and high rank total symmetry representation, 
%which are the dominant contributions to electroweak symmetry breaking.
Since the effects for the electroweak symmetry breaking by such type of the potential is known to be negligible, 
we do not consider them hereafter in this paper.  

A comment for giving a mass to exotic fermions is also given, 
In a model building in higher dimensional theory, exotic fermions cannot be avoid in general and have to be massive. 
It is easily done as follows. 
Introducing 4D fermions with opposite chirality and conjugated quantum numbers to the exotic fermions at the 4D fixed points, 
we can always give a mass to the exotic fermion through a Dirac mass term with such a 4D fermion. 

First, we consider the case of introducing fermions in the adjoint representation.
Since the differences are the degrees of freedom and an overall sign in the potential 
between the adjoint representation fermions and the gauge fields, the effective potential is totally given by
	\begin{align}
		V_{\text{eff}}^{\text{1-loop}} 
		&= V^g_{\text{eff}} + V^{f_\text{ad}}_{\text{eff}} \notag \\
%		&=  -\frac{3 \cdot 3}{64 \pi^6 R^4}  \sum_{ k = 1}^\infty \left[\{ 1 + 2(-1)^k \}\nu(\alpha^{\prime}) + 2\{ 1 + (-1)^k \}\nu(\alpha^{\prime}/2) \right] \notag \\
%		&=-\frac{9}{64 \pi^6 R^4}  \sum_{ k = 1}^\infty \frac{1}{k^5}\left[ \{ 1 + 2(-1)^k \}\cos( 2 \pi k \alpha^\prime) + 2\{1 + (-1)^k \} \cos( \pi k \alpha^\prime)\right] \\
		&=\frac{3(4N_{\text{ad}}-3)}{64 \pi^6 R^4}  \sum_{ k = 1}^\infty  \frac{1}{k^5}\qty[\{ 1 + 2(-1)^k \}\cos( 2 \pi k \alpha)
 + 2\{1 + (-1)^k \}  \cos( \pi k \alpha)],
	\end{align}
where $N_{\text{ad}}$ is the number of the adjoint representation fermion.
The sign of the curvature of the potential at $\alpha_{\text{min}}$ depends on the value of $N_{\text{ad}}$ as follows.
\begin{align}
(N_{\text{ad}} = 0) \qquad 
% \sum_{l} l^2 a_l < \sum_{l} \frac{3}{4}l^2 b_l\quad \rightarrow 
 &\eval{\pdv{^2}{{\alpha}^2} V_{\text{eff}}^{\text{1-loop}}}_{\alpha=0} < 0,\quad 
 \eval{\pdv{^2}{{\alpha}^2}V_{\text{eff}}^{\text{1-loop}}}_{\alpha=\frac{1}{2}} > 0 \\
(N_{\text{ad}} > 0), \qquad 
%\sum_{l} l^2 a_l > \sum_{l} \frac{3}{4}l^2 b_l \quad \rightarrow
 &\eval{\pdv{^2}{{\alpha}^2}V_{\text{eff}}^{\text{1-loop}}}_{\alpha=0} > 0 , 
 \quad \eval{\pdv{^2}{{\alpha}^2} V_{\text{eff}}^{\text{1-loop}}}_{\alpha=\frac{1}{2}} < 0.
\end{align}
Therefore, the electroweak symmetry cannot be broken when only adjoint representation fermions are introduced.

\begin{figure}[H]
		\centering
		\includegraphics[width=100mm]{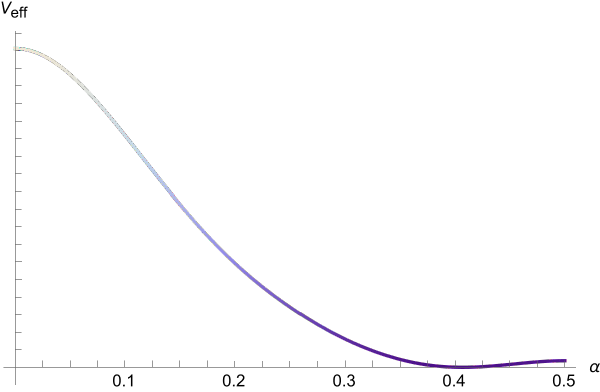}
		\caption{Effective potential of the gauge field and a 4-rank totally symmetric representation 
		$V^g_\text{eff} + V^{\text{$f_4$}}_{\text{eff}}$.
		The potential minimum is found to be $\alpha_{\text{min}}\simeq 0.4069.$
		%\quad ($\alpha_{\text{min}}<\frac{1}{2}$)
		}
		\label{fig:potentialplot4}
	\end{figure}

Next, we consider fermions in a  $Sp(6)$ fundamental representation and its KK mass spectrum.
%	\begin{equation}
%		\Psi =
%			\begin{pmatrix}
%				\mqty{\psi_1 \\ \psi_2\\ \psi_3\\ \psi_4 \\ \psi_5 \\ \psi_6}
%			\end{pmatrix}
%	\end{equation}
From the $Z_2$ invariance of $\mathcal{L}_5$, the boundary conditions of 5D fermion 
in the representation $R$ at the fixed points are 
	\begin{equation}
		\Psi(x^\mu, x^5_i-x^5) = R(P_i) (-i \Gamma^5 )\Psi(x^\mu, x^5_i+x^5).
	\end{equation}
The $Z_2$ parity of each component in the $Sp(6)$ sextet is given as
	\begin{align}
		\Psi = 
			\begin{pmatrix}
				\mqty{\psi_{1}^{(++)} \\[1mm] \psi_{2L}^{(++)} \\[1mm] \psi_3^{(-+)} \\ \hline 
				\psi_4^{(+-)} \\[1mm] \psi_5^{(+-)} \\[1mm] \psi_6^{(--)}}
			\end{pmatrix}.
			\label{sextetZ2parity}
	\end{align}
The $Sp(6)$ sextet can be decomposed into the $SU(3)$ representation: 
$\bf{6}\rightarrow3+\overline{3}$ and furthermore into the $SU(2)$ representation through $\bf{3} \rightarrow \bf{2} + \bf{1}$.
	\begin{align}
\bf{6}
%		\ytableausetup{mathmode, boxsize=2em}
%			\begin{ytableau}
%				6 
%			\end{ytableau}
		&\xRightarrow{SU(3)}
\
{\bf{3}+\overline{\bf{3}}}
\
%			\begin{ytableau}
%				3
%			\end{ytableau}%_{\tiny{\begin{array}{l} (++)\\(++)\\(-+)\end{array}}}
%		\quad + \quad
%		\overline{
%			\begin{ytableau}
%				3
%			\end{ytableau}}%_{\tiny{\begin{array}{l} (+-)\\(+-)\\(--)\end{array}}} 
%\notag \\[3mm]
%		&\quad = \quad 
%			\begin{ytableau}
%				{}
%			\end{ytableau}_{\tiny{\begin{array}{l} (++)\\(++)\\(-+)\end{array}}}
%		\quad  + \quad
%			\begin{ytableau}
%				*(lime) \\ *(lime)
%			\end{ytableau}_{\tiny{\begin{array}{l} (+-)\\(+-)\\(--)\end{array}}}\notag \\[7mm]
\xRightarrow{SU(2)}
\
{\bf{2}}_{(++)}
%			\begin{ytableau}
%				2
%			\end{ytableau}_{(++)}
		+
			 \bf{1}_{(-+)}
%		\quad  + \quad
+{\bf{2}}_{(+-)}
%		\overline{
%			\begin{ytableau}
%				%*(lime)
%				2
%			\end{ytableau}}_{(+-)}
		+
			\bf{1}_{(--)}. 
	\end{align}
From the above information of decomposition, the KK mass spectrum of $Sp(6)$ sextet can be read as follows. 
For the $SU(2)$ doublet, the KK index receives a constant shift as $n \to n+\frac{\alpha}{2}$. 
%Since only the $SU(2)$ doublet field interacts with the Higgs field, 
%the KK spectrum of the $SU(2)$ doublet fermion in the sextet of $Sp(6)$ is essentially 
%the same as the $SU(2)$ doublet component of the $Sp(6)$ gauge field.
%Thus, the fermion's one-loop effective potential can be computed using only the information from the $SU(2)$ decomposition.
Taking into account $Z_2$ parity, the KK spectrum of the $Sp(6)$ sextet can be found, 
	\begin{align}
		M_n^2 = 
		\left( \frac{n + \frac{\alpha }{2}}{R} \right)^2,\; 
		\left( \frac{n + \frac{1}{2}}{R} \right)^2,\;
		\left( \frac{(n + \frac{1}{2}) + \frac{\alpha }{2}}{R} \right)^2, \;
		\left(\; \frac{n }{R}\; \right)^2 .
	\end{align}
Since the KK mass spectrum of the $SU(2)$ singlet component is independent of the Higgs VEV parameter $\alpha$, 
the contribution to the potential can be neglected in our analysis of electroweak symmetry breaking. 
%but only gives a constant term in the effective potential.
This argument can be easily extended to a $N$-rank totally symmetric representation of $SU(2)$ doublets.
%To consider the covariant derivative of the $SU(2)$ $N$-rank totally symmetric representation, 
%consider a unitary transformation of fermions with symmetric subscripts $i_1, \cdots ,i_N$.
%From the symmetric subscripts, the number of subscripts corresponds to the coefficients over the gauge field.
%The covariant derivative of the $SU(2)$ $N$-rank totally symmetric representation is as follows
In this case, we need to know the covariant derivative of the $SU(2)$ $N$-rank totally symmetric representation, 
which is given by
\begin{align}
D_5\Psi_{i_1 \cdots i_N} = (\delta_{i_1i_1^\prime}\partial_5 + N i g A_5^a T^a_{i_1i_1^\prime})\Psi_{i_1^\prime \cdots i_N} .
\end{align}
where an additional factor $N$ appears from the symmetric indices $i_1, \cdots ,i_N$. 
This allows us to know the KK mass spectrum of the general $N$-rank totally symmetric representation fermions.
The $SU(2)$ $N$-rank totally symmetric representation is [${\bf{N+1}}$] dimensional representation of $SU(2)$.
The KK mass spectrum of the [${\bf{N+1}}$] dimensional representation with $Z_2$ parity $(++)$ is obtained below,
\begin{align}
(N= \text{odd}) \qquad &M^2_n = \left( \frac{n + \frac{N \alpha}{2}}{R} \right)^2,\ 
\left( \frac{n + \frac{(N-2) \alpha}{2}}{R} \right)^2, \cdots ,\left( \frac{n + \frac{\alpha}{2}}{R}\right)^2 ,\\[3mm]
(N= \text{even}) \qquad &M^2_n = \left( \frac{n + \frac{N \alpha}{2}}{R} \right)^2,\ 
\left( \frac{n + \frac{(N-2) \alpha}{2}}{R} \right)^2, \cdots ,\left( \frac{n + \alpha}{R}\right)^2.
\end{align} 
As will be seen below, the realistic electroweak symmetry breaking is possible in the $N=4$ case.
The 4-rank totally symmetric representation of $Sp(6)$ is ${\bf{126}}$ dimensional representation\footnote{More precisely, 
${\bf 126}$ representation considered in this paper is a ${\bf 126'}$ listed as Dynkin label (400) in \cite{Yamatsu, FKS}.}, 
which can be decomposed into the $SU(3)$ representation as follows\footnote{More precisely, ${\bf 15}$ in this paper is 
a ${\bf 15'}$ in \cite{Yamatsu, FKS}.}. 
\begin{equation}\label{126}
{\bf{126}} \quad \xRightarrow{SU(3)} \quad {\bf{27}} + \bf{24} + \overline{\bf{24}} + {\bf{15}} +\overline{\bf{15}}  +\bf{8}+\bf{6}+\overline{\bf{6}} + \bf{1}
\end{equation}
Each $SU(3)$ representation is further decomposed into $SU(2)$ representation with the corresponding $Z_2$ parity as follows. 
\begin{align}
	{\bf{27}} &\xRightarrow{SU(2)} {\bf{5}}_{(++)}\ +\ {\bf{4}}_{(-+)}\times 2\ 
	+\ {\bf{3}}_{(++)}\times 3 \ +\ {\bf{2}}_{(-+)}\times 2\ +\ {\bf{1}}_{(++)}, \notag \\
\bf{24},\overline{\bf{24}}  &\xRightarrow{SU(2)} {\bf{5}}_{(+-)}\ +\ {\bf{4}}_{(--)}\times 2\ 
+\ {\bf{3}}_{(+-)}\times 2 \ +\ {\bf{2}}_{(--)}\times 2\ +\ {\bf{1}}_{(+-)}, \notag \\
%\overline{\bf{15}} &\xRightarrow{SU(2)} {\bf{5}}_{(++)}+{\bf{4}}_{(-+)}+{\bf{3}}_{(++)} +{\bf{2}}_{(-+)}+{\bf{1}}_{(++)} \notag \\
\bf{15},\overline{\bf{15}}  &\xRightarrow{SU(2)} {\bf{5}}_{(++)}\ 
+\ {\bf{4}}_{(-+)}\ +\ {\bf{3}}_{(++)} \ +\ {\bf{2}}_{(-+)}\ +\ {\bf{1}}_{(++)}, \notag \\
\bf{8} &\xRightarrow{SU(2)} {\bf{3}}_{(+-)} \ +\ {\bf{2}}_{(--)}\times 2\ +\ {\bf{1}}_{(+-)}, \notag \\
\bf{6},\overline{\bf{6}} &\xRightarrow{SU(2)} {\bf{3}}_{(++)} \ +\ {\bf{2}}_{(-+)}\ +\ {\bf{1}}_{(++)}. 
\label{deco}
\end{align} 
%This situation is better put into perspective when represented by a Young tableau in Apendix B.
These decomposition can be easier to obtain by using the Young tableau as shown in Appendix B. 
Then, we obtain the effective potential of 4-rank totally symmetric representation of $Sp(6)$ as follows.\footnote{Note that 
the $\bf{4}$ representation has potential terms like $\cos(3\pi k \alpha)$ and $\cos(\pi k \alpha)$. 
Similarly, the $\bf{5}$ representation has those like $\cos(4\pi k \alpha)$ and $\cos(2\pi k \alpha)$. 
In reading off the potential from (\ref{deco}), this property should be considered.}
\begin{align}
V_{\text{eff}}^{f_4}
%&=  \frac{3 \cdot 4}{128 \pi^6 R^4}  \sum_{k \neq 0} \{( 3+ 2(-1)^k )\nu(2 \alpha^{\prime}) +( 4+ 4(-1)^k )\nu(3 \alpha^{\prime}/2) \notag \\
%& \hspace*{70pt} + ( 10 + 7(-1)^k )\nu(\alpha^{\prime}) + ( 10 + 10(-1)^k )\nu(\alpha^{\prime}/2) \} \notag \\[7mm]
&=\frac{3}{16 \pi^6 R^4}  \sum_{k = 1}^\infty  \frac{1}{k^5} [ \{3+ 2(-1)^k \} \cos( 4 \pi k \alpha)+\{ 4+ 4(-1)^k \}\cos( 3 \pi k \alpha) \notag \\
& \hspace*{85pt}+ \{10 + 7(-1)^k \} \cos( 2 \pi k \alpha) + \{ 10 + 10(-1)^k \} \cos( \pi k \alpha) ].
\end{align}
We analyze the effective potential from the gauge field and a fermion in the 4-rank totally symmetric representation 
$V_{\text{eff}}^{\text{1-loop}}=V_{\text{eff}}^g + V_{\text{eff}}^{f_4}$.
It has a negative curvature at $\alpha=0,\ \frac{1}{2}$:
\begin{align}
%\eval{\pdv{^2}{{\alpha}^2}V_{\text{eff}}^{f_4}}_{\alpha=0}
%=-\frac{327}{32 \pi^4 R^4} \\
%\eval{\pdv{^2}{{\alpha}^2}V_{\text{eff}}^{f_4}}_{\alpha=\frac{1}{2}} 
%=-\frac{321}{128 \pi^4 R^4}\\
%\sum_{l} l^2 a_l > \sum_{l} \frac{3}{4}l^2 b_l \quad \rightarrow
\eval{\pdv{^2}{{\alpha}^2}V_{\text{eff}}^{\text{1-loop}}}_{\alpha=0}
=&-\frac{75}{8 \pi^4 R^4} \zeta(3)\quad < 0 , \notag \\[3mm]
\eval{\pdv{^2}{{\alpha}^2}V_{\text{eff}}^{\text{1-loop}}}_{\alpha=\frac{1}{2}} 
=&-\frac{951}{512 \pi^4 R^4}  \zeta(3)
< 0.
\end{align}
The minimum is located in a range of $0<\alpha_{\text{min}}<1/2$ and 
the Figure \ref{fig:potentialplot4} shows $\alpha_{\text{min}}\simeq0.4069$.
Therefore, the contribution from the 4-rank totally symmetric representation can realize the realistic electroweak symmetry breaking. 
%similar to the Standard Model:
\begin{align}
Sp(6) \xrightarrow{Z_2\: \text{orbifold}} SU(2)_L 
\times U(1)_Y \times U(1)_X \xrightarrow{\text{quantum effect}} U(1)_{em} \times U^\prime(1) \notag 
\end{align}

%%%%%%%%%%%%%%%%%%%%%%%%%%%%%%%%%%%%%%%%
\section{Renormalization group equation analysis of weak mixing angle}
%%%%%%%%%%%%%%%%%%%%%%%%%%%%%%%%%%%%%%%%
In this section, we study how our prediction for the weak mixing angle is good 
by one-loop renormalization group equation (RGE).  
One-loop RGE for the gauge couplings are  
\begin{align}
\frac{1}{g^2_i(\mu)}-\frac{1}{g^2_i(M_Z)} = \frac{b_i}{8\pi^2} \ln \left( \frac{M_Z}{\mu} \right), 
\end{align}
where the index $i=1,2,3$ labels the gauge coupling $g_{1,2,3}$, 
$b_i$ is the corresponding beta function coefficients and $\mu$ is a renormalization scale. 

The RGE between the compactification scale $1/R$ and the half of it $1/(2R)$, 
and between the half of the compactification scale and the Z-boson mass scale $M_Z$ are given as
\begin{align}
\frac{1}{g^2_i(1/R)}-\frac{1}{g^2_i(1/(2R))} &= \frac{1}{8\pi^2} (b_i^{{\rm SM}} + \Delta b_i) \ln \left( \frac{1}{2} \right), \\
\frac{1}{g^2_i(1/(2R))}-\frac{1}{g^2_i(M_Z)} &= \frac{1}{8\pi^2} b_i^{{\rm SM}} \ln \left( 2M_Z R \right), 
\end{align}
where $b_i^{{\rm SM}}$ is the beta function coefficient of the Standard Model 
and $\Delta b_i$ stands for the beta function coefficients by the contributions from the fields with mass $1/(2R)$. 
Eliminating $1/(2R)$, we obtain 
\begin{align}
\frac{1}{g^2_i(1/R)}-\frac{1}{g^2_i(M_Z)} &= \frac{1}{8\pi^2} (b_i^{{\rm SM}} \ln (M_Z R) - \Delta b_i \ln 2 ). 
\label{RGE1}
\end{align}
The weak mixing angle at the compactification scale can be rewritten into that at the $Z$-boson mass scale 
by using the RGE,  
\begin{align}
\frac{1}{4} = \sin^2 \theta_W  = \frac{g_Y^2}{g_2^2 + g_Y^2} 
= \frac{1}{1 + \frac{3g_2^2}{g_1^2}}
= \left[
1+3\frac{\frac{1}{g_1^2(M_Z)} + \frac{1}{8\pi^2}(b_1^{{\rm SM}} \ln (M_Z R) - \Delta b_1 \ln 2)}
{\frac{1}{g_2^2(M_Z)} + \frac{1}{8\pi^2}(b_2^{{\rm SM}} \ln (M_Z R) - \Delta b_2 \ln 2)}
\right]^{-1},
\end{align}
where a relation $g_Y=g_1/\sqrt{3}$ is put in the third equality and the RGE (\ref{RGE1}) is utilized in the last equality. 
After some algebra, we find an expression for the weak mixing angle at the $Z$-boson masse scale as
\begin{align}
\sin^2 \theta_W(M_Z) 
&= \left[ 4 
-\frac{3 g_2^2(M_z)}{8\pi^2} \left\{ (b_1^{{\rm SM}} - b_2^{{\rm SM}}) \ln \left( \frac{M_Z}{2M_W} \alpha \right)
+ \left( \Delta b_1 -\Delta b_2 \right)\ln 2\right \} \right]^{-1} 
\label{mixingatZ}
\end{align}
Here the compactification scale is replaced by the $M_W$ and $\alpha$ through the relation $M_W=\frac{\alpha}{2R}$.
%One of the remarkable features is that this expression is independent of the compactification scale. 

In order to obtain the weak mixing angle at the $Z$-boson mass scale, 
we have only to calculate the beta function coefficients $\Delta b_{1,2}$ from the contributions of the fields with mass $1/(2R)$. 
Such kind fields included in our model are the gauge fields, the scalar fields from the fifth component of the gauge field $A_5$, 
and fermions in the %adjoint and 
${\bf 126}$ dimensional representations of $Sp(6)$. 
Let us consider each contribution in order. 

As for the gauge field and the scalar fields of $A_5$, 
we first notice a decomposition of the $Sp(6)$ adjoint representation into $SU(3)$ and $SU(2)_L \times U(1)_Y$ ones, 
\begin{align}
{\bf 21} &\xRightarrow{SU(3)} {\bf 8} + {\bf 6} + {\bf \overline{6}} + {\bf 1} \nonumber \\
&\xRightarrow{SU(2)} ({\bf 3}_0 + {\bf 2}_{-3/2} + {\bf 2}_{3/2} + {\bf 1}_0) + ({\bf 3}_{-1} + {\bf 2}_{1/2} + {\bf 1}_2) 
+ ({\bf \overline{3}}_1 + {\bf \overline{2}}_{-1/2} + {\bf \overline{1}}_{-2}) + {\bf 1}_0. 
\end{align}
In our model, the $U(1)_Y$ hypercharge is given 
by an $SU(2)$ decomposition of ${\bf 6}$ representation 
${\bf 6}=({\bf 2}_{-1/2} + {\bf 1}_{1}) + (\overline{{\bf 2}}_{1/2} + {\bf \overline{1}}_{-1})$. 
Noticing that the fields with mass $1/(2R)$ have the $Z_2$ parities as $(+ -)$ or $(- +)$, 
we find such kind of the gauge fields from the $Z_2$ parity for the $Sp(6)$ sextet (\ref{sextetZ2parity}), 
${\bf  2}_{-3/2}, {\bf 2}_{3/2}$ from ${\bf 8}$, 
${\bf 3}_{-1}$ from ${\bf 6}$ and ${\bf \overline{3}}_{1}$ from ${\bf \bar{6}}$. 
Corresponding beta function coefficients are calculated as follows. 
\begin{align}
-\Delta b_1({\rm gauge}) &= b_1({\bf 3}_{-1}) + b_1({\bf 2}_{-3/2}) + b_1({\bf \overline{3}}_{1}) + b_1({\bf \overline{2}}_{3/2})  \notag \\[2mm]
&= \left( \frac{11}{3} - \frac{1}{6} \right) \left[ 3 \times 1^2 + 2 \times \left( \frac{3}{2} \right)^2 \right] \times 2 = \frac{105}{2}, \\[3mm]
-\Delta b_2({\rm gauge}) &= b_2({\bf 3}_{-1}) + b_2({\bf 2}_{-3/2}) + b_2({\bf \overline{3}}_{1}) + b_2({\bf \overline{2}}_{3/2}) \nonumber \\[2mm]   
&= \frac{11}{3} [2 C_2({\bf 3}) + 2 C_2({\bf 2})]
 - \frac{1}{6} [2 T_2({\bf 3}) + 2 T_2({\bf 2})] = \frac{58}{3}, 
\label{betagauge}
\end{align}
where $C_2({\bf{R}})$ and $T_2({\bf R})$ of the representation $\bf{R}$ are defined as follows.
$f^{abc}f^{dbc} = C_2({\bf R}) \delta^{ad}~(f^{abc}: {\rm structure~constant})$ and ${\rm Tr}(T^a({\bf{R}})T^b({\bf{R}}))=T_2({\bf{R}})\delta^{ab}$. 
Note that $C_2({\bf 3})=T_2({\bf \overline{3}})= 2, C_2({\bf 2}) = 3/4$, $T({\bf 2})=1/2, T({\bf 3})= 2\times 2T({\bf 2})= 2$.

%The beta function coefficients for the adjoint fermions are similarly calculated as
%\begin{align}
%-\Delta b_2({\rm adj~fermions}) &= b_2({\bf 3}_{-1}) + b_2({\bf 2}_{-3/2}) + b_2({\bf \overline{3}}_{1}) + b_2({\bf \overline{2}}_{3/2}) \nonumber \\   
%&= \frac{2}{3} [2 C_2({\bf 3}) + 2 C_2({\bf 2})] = \frac{10}{3}, \\
%-\Delta b_1({\rm adj~fermions}) &= b_1({\bf 3}_{-1}) + b_1({\bf 2}_{-3/2}) + b_1({\bf \overline{3}}_{1}) + b_1({\bf \overline{2}}_{3/2}) \nonumber \\
%&= \frac{2}{3} \left[ 3 \times 1^2 + 2 \times \left( \frac{3}{2} \right)^2 \right] \times 2 = 10. 
%\label{betaadj}
%\end{align}

As for the fermion in the ${\bf 126}$ dimensional representation, the decomposition into $SU(3)$ representations is given by \eqref{126}
%\begin{align}
%{\bf 126} \xRightarrow{SU(3)} {\bf 27} + {\bf 24} + {\bf \overline{24}} + {\bf 15} + {\bf \overline{15}} + {\bf 8} + {\bf 6} + {\bf \overline{6}} + {\bf 1}
%\end{align}
and the fields of mass $1/(2R)$ with $Z_2$ parities $(+-), (-+)$ in $SU(2)_L \times U(1)_Y$ representation are found as 
\begin{align}
{\bf 27} &\supset {\bf 4}_{-3/2} + {\bf \overline 4}_{3/2} + {\bf 2}_{-3/2} + {\bf 2}_{3/2}, \nonumber \\
{\bf 24} &\supset {\bf 5}_{-1} + {\bf 3}_{-1} + {\bf 3}_{2}, \quad %\nonumber \\
{\bf \overline{24}} \supset {\bf \overline{5}}_{1} + {\bf \overline{3}}_{1} + {\bf \overline{3}}_{-2}, \nonumber \\
{\bf 15} &\supset {\bf 4}_{-1/2} + {\bf 2}_{5/2}, \qquad %\nonumber \\
{\bf \overline{15}} \supset {\bf \overline{4}}_{1/2} + {\bf 2}_{-5/2}, \nonumber \\
{\bf 8} &\supset {\bf 3}_0, \quad%\nonumber \\
{\bf 6} \supset {\bf 2}_{-1/2}, \quad %\nonumber \\
{\bf \overline{6}} \supset {\bf 2}_{1/2}.
\end{align}
Corresponding beta function coefficients are calculated as follows. 
\begin{align}
-\Delta b_1({\bf 126}) &= 
b_1({\bf 2}_{-1/2}) + b_1({\bf 2}_{1/2}) + b_1({\bf 3}_0)
+ b_1(\overline{{\bf 4}}_{1/2}) + b_1({\bf 2}_{-5/2})
+ b_1({\bf 4}_{-1/2}) + b_1({\bf 2}_{5/2}) \nonumber \\
&\quad+ b_1(\overline{{\bf 5}}_{1}) + b_1( \overline{{\bf 3}}_{1}) + b_1(\overline{{\bf 3}}_{-2}) + b_1({\bf 5}_{-1}) + b_1({\bf 3}_{-1}) + b_1({\bf 3}_{2}) 
\nonumber \\
&\qquad+ b_1({\bf \overline 4}_{3/2}) + b_1({\bf 4}_{-3/2}) + b_1({\bf 2}_{3/2}) + b_1({\bf 2}_{-3/2}) \nonumber \\[2mm]
&= -\frac{4}{3} \left[ 
\left( \frac{1}{2} \right)^2 \times 2 \times 2 + \left( \frac{1}{2} \right)^2 \times 4 \times 2  
+ \left( \frac{5}{2} \right)^2 \times 2 \times 2 \right. \nonumber \\
& \left. 
\qquad+ (1^2 \times 5 + 1^2 \times 3 + 2^2 \times 3) \times 2 
+ \left( \frac{3}{2} \right)^2 \times 4 \times 2
+ \left( \frac{3}{2} \right)^2 \times 2 \times 2 
\right] \notag \\
&= -\frac{380}{3}, \\[5mm]
-\Delta b_2({\bf 126})  &= 
b_2({\bf 2}_{-1/2}) + b_2({\bf 2}_{1/2}) + b_2({\bf 3}_0)
+ b_2(\overline{{\bf 4}}_{1/2}) + b_2({\bf 2}_{-5/2})
+ b_2({\bf 4}_{-1/2}) + b_2({\bf 2}_{5/2}) \nonumber \\
&\quad+ b_2(\overline{{\bf 5}}_{1}) + b_2( \overline{{\bf 3}}_{1}) + b_2(\overline{{\bf 3}}_{-2}) + b_2({\bf 5}_{-1}) + b_2({\bf 3}_{-1}) + b_2({\bf 3}_{2}) 
\nonumber \\
&\qquad+ b_2({\bf \overline 4}_{3/2}) + b_2({\bf 4}_{-3/2}) + b_2({\bf 2}_{3/2}) + b_2({\bf 2}_{-3/2}) \nonumber \\[2mm]
&= -\frac{4}{3} \left[ 6C_2({\bf 2}) + 5C_2({\bf 3}) + 4C_2({\bf 4}) + 2C_2({\bf 5}) \right]
= -\frac{212}{3}, 
\label{beta126}
\end{align}
where the relations for the quadratic Dynkin indices $C_2({\bf 4}) = \frac{(2+2)(2+3)}{2}C_2({\bf 2})=5$, 
$C_2({\bf 5}) = \frac{(2+2)(2+3)(2+4)}{6}C_2({\bf 2})=10$ are applied. 
Collecting all of the results (\ref{betagauge}), %(\ref{betaadj}) 
and (\ref{beta126}) together, 
we obtain the beta function coefficients contributing from the fields with mass $1/(2R)$
\begin{align}
\Delta b_1 = \frac{445}{6}, \qquad \Delta b_2 =  \frac{154}{3}. 
\end{align}

Substituting these results, $\frac{4\pi}{g_2^2(M_Z)} = 29.57 \dots$, $b_1^{{\rm SM}}=\frac{41}{6}, b_2^{{\rm SM}}=-\frac{19}{6}$ and 
$\alpha =0.4069$ (Fig. 2) into (\ref{mixingatZ}), 
we find the weak mixing angle at the $Z$-boson mass scale 
\begin{align}
\sin^2 \theta_W(M_Z) 
&= \left[ 4 
-\frac{3 g_2^2(M_z)}{8\pi^2} \left\{ (b_1^{{\rm SM}} - b_2^{{\rm SM}}) \ln \left( \frac{M_Z}{2M_W} \alpha \right)
%+ \left( \frac{40}{3} - \frac{56}{3} \right)
+\frac{133}{6}\ln 2\right \} \right]^{-1} 
%\nonumber \\[2mm]
%&\simeq (4 - 0.009)^{-1} &
\sim 0.2501. 
\end{align}
%The result is remarkably good agreement with the experimental data $\sin^2 \theta_W \sim 0.23$ at weak scale. 

To improve this result, we introduce fermions in the adjoint representation of $Sp(6)$. 
We extract only fermions with a mass $1/(2R)$ in the the nontrivial $SU(2)$ representations from 
%The adjoint representation of $Sp(6)$ is 
the decomposition under $SU(2) \times U(1)$.
\begin{align}
 {\bf{21}} \supset %\rightarrow
%\quad {\bf{3}}_0 + 
{\bf{2}}_{\frac32} +{\bf{2}}_{-\frac32} %+ {\bf{1}}_0 \notag \\
 +  {\bf{3}}_{-1} %+ {\bf{2}}_{-\frac12} %+ {\bf{1}}_{-2}\notag \\
 + \overline{\bf{3}}_{1} %+ {\bf{2}}_{\frac12} %+ {\bf{1}}_{2}\notag \\
%& + {\bf{1}}_0
%&\quad {\bf{3}}_0 + {\bf{2}}_{\frac32} + \overline{\bf{2}}_{-\frac32} + {\bf{1}}_0 \notag \\
%& +  {\bf{3}}_1 + {\bf{2}}_{-\frac12} + {\bf{1}}_{-2}\notag \\
%& + {\bf{3}}_{-1} + {\bf{2}}_{\frac12} + {\bf{1}}_{2}\notag \\
%& + {\bf{1}}_0
\end{align}
Corresponding beta function coefficients are calculated as follows. 
\begin{align}
-\Delta b_1({\bf 21}) &= 
b_1({\bf 2}_{3/2}) + b_1({\bf 2}_{-3/2}) + b_1(\overline{\bf 3}_{1}) + b_1({\bf 3}_{-1}) \nonumber \\
%&\quad+ b_1({\bf 5}_{1}) + b_1({\bf 3}_{1}) + b_1({\bf 3}_{-2}) + b_1({\bf 5}_{-1}) + b_1({\bf 3}_{-1}) + b_1({\bf 3}_{2}) 
%\nonumber \\
%&\qquad+ b_1({\bf 4}_{3/2}) + b_1({\bf 4}_{-3/2}) + b_1({\bf 2}_{3/2}) + b_1({\bf 2}_{-3/2}) \nonumber \\[2mm]
&= -\frac{4}{3} \left[ 
%\left( 
1^2 \times 3 + \left( \frac{3}{2} \right)^2 \times 2 \right]\times 2 %\nonumber \\&
= -20, \\[5mm]
-\Delta b_2({\bf 21}) &= 
b_1({\bf 2}_{3/2}) + b_1({\bf 2}_{-3/2}) + b_1(\overline{\bf 3}_{1}) + b_1({\bf 3}_{-1}) \nonumber \\
&= -\frac{4}{3} \left[ 2C_2({\bf 2}) + 2C_2({\bf 3}) \right]
= -\frac{20}{3}. 
\label{beta21}
\end{align}
In the case where the $N_{{\rm ad}}$ adjoint fermions are introduced, $\Delta_{1,2}$ are modified as follows. 
\begin{align}
\Delta b_1 = \frac{445}{6} + 20N_{{\rm ad}}, \quad 
\Delta b_2 = \frac{154}{3} + \frac{20}{3} N_{{\rm ad}}. 
\end{align}
The VEV $\alpha$ obtained by the re-analysis of the potential minimization 
and the weak mixing angle at the weak scale evaluated by (\ref{mixingatZ}) are summarized in Tabe \ref{result}.   
We find that the weak mixing angle at the weak scale can be in good agreement with the experimental data. 
%\begin{comment}
\begin{table}[h]
\begin{center}
\begin{tabular}{|c|c|c|}
\hline
$N_{\text{ad}}$ & $\alpha$ & $\sin^2 \theta_W(M_Z)$ \\
\hline
1 & 0.370 & 0.242 \\
2 & 0.327 & 0.235 \\
3 & 0.285 & 0.228 \\
4 & 0.251 & 0.221 \\
5 & 0.226 & 0.215 \\
\hline
\end{tabular}
\caption{Sample points of VEV of Higgs field and the weak mixing angle at the weak scale for the number of the adjoint fermions.}
\label{result}
\end{center}
\end{table}
%\end{comment}

%%%%%%%%%%%%%%
\section{Conclusion}
%%%%%%%%%%%%%
In this paper, we have studied whether the electroweak symmetry breaking will happen 
in a five dimensional $Sp(6)$ Gauge-Higgs Unification model.  
One of the advantages to consider this model is 
that the weak mixing angle is predicted to be $\sin^2 \theta_W = 1/4$ at the compactification scale \cite{HLM}. 
In GHU, the non-local Higgs potential is generated at one-loop level 
although the Higgs potential at tree level is vanished by the gauge invariance. 
%Therefore, Higgs mass is likely to be small and obtaining a realistic Higgs mass is a very non-trivial problem in GHU. 
In our analysis, it is shown that the realistic electroweak symmetry breaking can be realized 
by introducing an additional fermion in a 4-rank totally symmetric representation.  %and thirty one adjoint representations.  
Furthermore, we have calculated the weak mixing angle at the weak scale by using the 1-loop RGE of the gauge coupling constants. 
Taking into account the contributions of the fields with a mass of $1/(2R)$ to the 1-loop RGE, 
it was found that the weak mixing angle predicted to be $\sin^2 \theta_W =1/4$ at the compactification scale 
%is almost preserved at the weak scale. 
can be corrected to be $\sin^2\theta_W \sim 0.23$ at the weak scale by introducing some fermions in the adjoint representation, 
which is in good agreement with the experimental data. %$\sin^2 \theta_W \sim 0.23$. 
 
A comment on Higgs mass is given.  
In our $Sp(6)$ GHU model with a fermion in a 4-rank totally symmetric representation, 
we find the Higgs mass to be around 13 GeV, which is too small.   
One of the approaches to improve this result has been known to utilize the gauge kinetic terms localized at the fixed points\cite{SSS}. 
Their effects is known to enhance the magnitude of the compactification scale, 
which makes a viable Higgs mass easier. 
If we introduce the gauge kinetic terms localized at the fixed points as
\begin{align}
{\cal L} = -\frac{1}{4g_5^2} F_{MN} F^{MN} 
-\delta(y) (2\pi R c_1) \frac{1}{4g_1^2} F_{\mu\nu} F^{\mu\nu} 
- \delta(y-\pi R) (2\pi R c_2) \frac{1}{4g_2^2} F_{\mu\nu} F^{\mu\nu},
\end{align}
where $c_{1,2}$ are arbitrary dimensionless constants. 
The second and the third terms are the 4D gauge kinetic terms with 4D gauge couplings $g_{1,2}$ 
localized at the fixed points $y=0, \pi R$, respectively. 
As discussed in \cite{MY}, 
one of the effects from these terms is an enhancement of the compactification as $1/R \to \sqrt{1+c_1+c_2}/R$, 
which implies that the Higgs mass of 125 GeV can be obtained by adjusting $c_1+c_2 \sim 90$.  
Furthermore, we expect that the weak mixing angle does not change drastically 
for a few order enhancement of the compactification scale 
since the dependence of the compactification scale on the weak mixing angle is logarithmic. 
More precisely, we have to study carefully the effects on the KK mass spectrum of the gauge field and the potential, 
but the required analysis is more complicated and beyond the scope of this paper. 
This issue is left for our future work. 

%%%%%%%%%%%%
\subsection*{Acknowledgments}
%%%%%%%%%%%
This work was supported by JST SPRING, Grant Number JPMJSP2139 (A.N.).

%\newpage
%%%%%%%%%%%%%%%%%%%%%%%%%%%%%%%%%%%%%%%%%%
\appendix
%%%%%%%%%%%%%%%%%%%%%%%%%%%%%%%%%%%%%%%%%
\section{$Sp(6)$ generators}
%%%%%%%%%%%%%%%%%%%%%%%%%%%%%%%%%%%%%%%%
The $Sp(6)$ generators $T^a\ (a=1, \cdots, 21)$ are given by $T^a=\frac{1}{2\sqrt{2}}t^a$ with the following $t^a$.
	\begin{align}
		t^a &=
			\begin{pmatrix}\mqty{
				\lambda^a & 0 \\
				0 & -(\lambda^a)^*
			}\end{pmatrix} \qquad (a = 1, \cdots 8),\\[3mm]
		t^9 &= \sqrt{\frac{2}{3}}
			\begin{pmatrix}\mqty{
				I & 0 \\
				0 & -I
			}\end{pmatrix} ,\\[3mm]
		t^a &= 
			\begin{pmatrix}\mqty{
				0 & M_j \\
				M_j & 0
			}\end{pmatrix} \qquad (a = 9+j,\; j=1 \cdots 6),\\[3mm]
		t^a &= 
			\begin{pmatrix}\mqty{
				0 & -iM_j \\
				iM_j & 0
			}\end{pmatrix} \qquad (a = 15+j,\; j=1 \cdots 6),
	\end{align}
where $\lambda^a$ is a Gell-Mann matrices and $M_j$ is
	\begin{align}
		M_1 &= 
			\begin{pmatrix}\mqty{
				\sqrt{2} & 0 & 0\\
				0 & 0 & 0\\
				0 & 0 & 0
			}\end{pmatrix} ,\;
		M_2 = 
			\begin{pmatrix}\mqty{
				0 & 0 & 0\\
				0 & \sqrt{2} & 0\\
				0 & 0 & 0
			}\end{pmatrix}  ,\;
		M_3 = 
			\begin{pmatrix}\mqty{
				0 & 0 & 0\\
				0 & 0 & 0\\
				0 & 0 & \sqrt{2}
			}\end{pmatrix} ,\notag \\[3mm]
		M_4 &= 
			\begin{pmatrix}\mqty{
				0 & 1 & 0\\
				1 & 0 & 0\\
				0 & 0 & 0
			}\end{pmatrix} ,\;
		M_5 = 
			\begin{pmatrix}\mqty{
				0 & 0 & 0\\
				0 & 0 & 1\\
				0 & 1 & 0
			}\end{pmatrix}  ,\;
		M_6 = 
			\begin{pmatrix}\mqty{
				0 & 0 & 1\\
				0 & 0 & 0\\
				1 & 0 & 0
			}\end{pmatrix} .
	\end{align}
These $T^a$ are normalized as $\Tr(T^a T^b)=\frac{1}{2}\delta^{ab}$.
\section{Decomposition of 4-rank totally symmetric representation}
%%%%%%%%%%%%%%%%%%%%%%%%%%%%%%%%%%%%
In this appendix B, we summarize the decomposition of the 4-rank totally symmetric representation of $Sp(6)$ 
into $SU(2)$ irreducible representations by using Young tablaeu. 

First, the decomposition of $Sp(6)$ sextet into $SU(2)$ irreducible representations ${\bf{6}} \xRightarrow{SU(3)} {\bf{3}} + \overline{\bf{3}} \xRightarrow{SU(2)} {\bf{2}}+{\bf{1}} + \overline{\bf{2}} + {\bf{1}}$ is shown.% by using Young tablaeu. 

	\begin{align}
		\ytableausetup{mathmode, boxsize=2em}
			\begin{ytableau}
				6
			\end{ytableau}%_{\tiny{\begin{array}{l} (++)\\(++)\\(-+)\end{array}}}
		&\xRightarrow{SU(3)}
%			\begin{ytableau}
%				3
%			\end{ytableau}%_{\tiny{\begin{array}{l} (++)\\(++)\\(-+)\end{array}}}
%		\quad + \quad
%		\overline{
%			\begin{ytableau}
%				3
%			\end{ytableau}}%_{\tiny{\begin{array}{l} (+-)\\(+-)\\(--)\end{array}}} 
%\notag \\
%		&\quad = \quad 
			\begin{ytableau}
				{}
			\end{ytableau}%_{\tiny{\begin{array}{l} (++)\\(++)\\(-+)\end{array}}}
		\quad  + \quad
			\begin{ytableau}
				*(lime) \\ *(lime)
			\end{ytableau}%_{\tiny{\begin{array}{l} (+-)\\(+-)\\(--)\end{array}}}
\notag \\
		&\xRightarrow{SU(2)}
			\begin{ytableau}
				{}
			\end{ytableau}_{(++)}
		+
			 \bf{1}_{(-+)}
		\quad  + \quad
		\overline{
			\begin{ytableau}
				*(lime){}
			\end{ytableau}}_{(+-)}
		+
			\bf{1}_{(--)}
	\end{align}
Here, the ${\bf{2}}$ and $\overline{\bf{2}}$ representation are essentially the same, but the ${\bf{2}}$ obtained by decomposing the ${\bf{3}}$ and  the $\overline{\bf{2}}$ obtained by decomposing the $\overline{\bf{3}}$ are different in parity sign
In order to distinguish between them, $\overline{\bf{2}}$ is marked with a colored box.
Hereafter $\otimes$ means symmetric product.

Next, we look at the decomposition of  the 4-rank totally representation of $Sp(6)$.
\begin{align}
\ytableausetup{mathmode, boxsize=2em}
& \ydiagram[ ]{4}  \notag \\[7mm]
%&\quad = \quad \ydiagram[  ]{2} \otimes \ydiagram[  ]{2} \notag \\[7mm]
&\xRightarrow{SU(3)} \left( \;  \ydiagram[ *(lime) ]{2, 2} + \ydiagram[ *(lime) ]{1, 1}*[ *(white) ]{1+1} 
+ \ydiagram[]{2} + {\bf{1}} \; \right) \otimes \left( \;   \ydiagram[ *(lime) ]{2, 2} + \ydiagram[  *(lime)  ]{1, 1}*[  *(white)  ]{1+1} 
+  \ydiagram[]{2} +{\bf{1}} \; \right) \notag \\[7mm]
&\quad = \quad \ydiagram[ *(lime) ]{2, 2}*[*(white)]{2+2} + \ydiagram[ *(lime) ]{1,1}*[*(white)]{1+3} 
+\ydiagram[ *(lime) ]{3, 3}*[*(white)]{3+1} \notag \\
& \qquad + \quad \ydiagram[ *(lime) ]{4,4}+  \ydiagram[]{4} +\ydiagram[ *(lime) ]{1, 1}*[*(white)]{1+1}
+\ydiagram[ *(lime) ]{2, 2}+ \ydiagram[]{2}+ {\bf{1}}
\end{align}
%where the Young tablaeu in colored box means a $\bar{\bf{3}}$ in $SU(3)$. 
Therefore, we obtain the decomposition the 4-rank totally representation of $Sp(6)$ into $SU(3)$ irreducible representation: 
\begin{equation}
\bf{126} \quad = \quad {\bf{27}} + \bf{24} + \overline{\bf{24}} + {\bf{15}} +\overline{\bf{15}}  +\bf{8}+\bf{6}+\overline{\bf{6}} + \bf{1}. 
\end{equation} 
Then, the decompositions of each multiplet into $SU(2)$ irreducible representaitons are listed. 

\noindent
(i)\  {\bf{27}} rep. %$\sim \overline{\bf{6}}\otimes \bf{6}$
\begin{align}
 &\ydiagram[ *(lime) ]{2, 2}*[*(white)]{2+2} \notag \\[5mm]
% & \quad = \quad \ydiagram[*(lime)]{2, 2} \otimes \ydiagram[]{2} \notag \\[5mm]
 & \xRightarrow{SU(2)} \left( \; \ydiagram[*(lime)]{2}_{(++)} + \ydiagram[*(lime)]{1}_{(-+)} 
 +{\bf{1}}_{(++)}\; \right) \otimes \left( \; \ydiagram[]{2}_{(++)} + \ydiagram[]{1}_{(-+)} +{\bf{1}}_{(++)} \; \right) \notag \\[7mm]
 & \quad = \quad \ydiagram[*(lime)]{2}*[*(white)]{2+2}_{(++)} +\ydiagram[*(lime)]{1}*[*(white)]{1+2}_{(-+)}+ \ydiagram[]{2}_{(++)} \notag \\
& \quad \quad + \ydiagram[*(lime)]{2}*[*(white)]{2+1}_{(-+)} + \ydiagram[*(lime)]{1}*[*(white)]{1+1}_{(++)} + \ydiagram[]{1}_{(-+)}\notag \\
& \quad \quad + \ydiagram[*(lime)]{2}_{(++)} + \ydiagram[*(lime)]{1}_{(-+)} + {\bf{1}}_{(++)}
\end{align}
(ii)\ {\bf{24}} rep.
\begin{align}
 &\ydiagram[ *(lime) ]{1,1}*[*(white)]{1+3} \notag \\[5mm]
% & \quad = \quad \ydiagram[*(lime)]{1, 1}*[*(white)]{1+1} \otimes \ydiagram[]{2} \notag \\[5mm]
& \xRightarrow{SU(2)} \left( \; \ydiagram[*(lime)]{1}*[*(white)]{1+1}_{(+-)} +\ydiagram[]{1}_{(--)} 
+ \ydiagram[*(lime)]{1}_{(--)} +{\bf{1}}_{(+-)}\; \right)\notag \\[2mm]
& \hspace*{250pt} \otimes \left( \; \ydiagram[]{2}_{(++)} + \ydiagram[]{1}_{(-+)} +{\bf{1}}_{(++)} \; \right) \notag \\[5mm]
 & \quad = \quad \ydiagram[*(lime)]{1}*[*(white)]{1+3}_{(+-)} +\ydiagram[]{3}_{(--)}
 + \ydiagram[*(lime)]{1}*[*(white)]{1+2}_{(--)} + \ydiagram[]{2}_{(+-)} \notag \\
& \quad \quad + \ydiagram[*(lime)]{1}*[*(white)]{1+1}_{(+-)} + \ydiagram[]{1}_{(--)}\notag \\
& \quad \quad + \ydiagram[*(lime)]{1}_{(--)} + {\bf{1}}_{(+-)}
\end{align}
(iii)\  $\overline{{\bf{24}}}$ rep.
\begin{align}
&\ydiagram[ *(lime) ]{3, 3}*[*(white)]{3+1} \notag \\[5mm]
%& \quad = \quad \ydiagram[*(lime)]{2, 2} \otimes \ydiagram[*(lime)]{1, 1}*[*(white)]{1+1} \notag \\[5mm]
& \xRightarrow{SU(2)} \left( \; \ydiagram[*(lime)]{2}_{(++)} + \ydiagram[*(lime)]{1}_{(-+)} +{\bf{1}}_{(++)}\; \right)
%\notag \\[2mm] & \hspace*{100pt} 
\otimes \left( \; \ydiagram[*(lime)]{1}*[*(white)]{1+1}_{(+-)} +\ydiagram[]{1}_{(--)} 
+ \ydiagram[*(lime)]{1}_{(--)} +{\bf{1}}_{(+-)}\; \right) \notag \\[5mm]
& \quad = \quad \ydiagram[*(lime)]{3}*[*(white)]{3+1}_{(+-)} + \ydiagram[*(lime)]{2}*[*(white)]{2+1}_{(--)} 
+ \ydiagram[*(lime)]{1}*[*(white)]{1+1}_{(+-)} \notag \\
& \quad \quad  + \ydiagram[]{1}_{(--)}\notag \\
& \quad \quad + \ydiagram[*(lime)]{3}_{(--)} + \ydiagram[*(lime)]{2}_{(+-)} + \ydiagram[*(lime)]{1}_{(--)}\notag \\
& \quad \quad + {\bf{1}}_{(+-)}\notag \\
\end{align}
(iv)\  {\bf{15}} rep.
\begin{align}
& \ydiagram[]{4} \notag \\[5mm]
%& \quad = \quad \ydiagram[]{2} \otimes\ydiagram[]{2} \notag \\[5mm]
& \xRightarrow{SU(2)} \left( \; \ydiagram[]{2}_{(++)} + \ydiagram[]{1}_{(-+)} +{\bf{1}}_{(++)} \; \right)  
\otimes \left( \; \ydiagram[]{2}_{(++)} + \ydiagram[]{1}_{(-+)} +{\bf{1}}_{(++)} \; \right) \notag \\[7mm]
& \quad = \quad \ydiagram[]{4}_{(++)} + \ydiagram[]{3}_{(-+)} + \ydiagram[]{2}_{(++)} \notag \\
& \quad \quad +  \ydiagram[]{1}_{(-+)} \notag \\
&\quad \quad + {\bf{1}}_{(++)}
\end{align}
(v)\ $\overline{\bf{15}}$ rep.
\begin{align}
& \ydiagram[*(lime)]{4,4} \notag \\[5mm]
%& \quad = \quad \ydiagram[*(lime)]{2, 2} \otimes \ydiagram[*(lime)]{2, 2} \notag \\[5mm]
& \xRightarrow{SU(2)} \left( \; \ydiagram[*(lime)]{2}_{(++)} + \ydiagram[*(lime)]{1}_{(-+)} +{\bf{1}}_{(++)}\; \right) 
\otimes  \left( \;  \ydiagram[*(lime)]{2}_{(++)} + \ydiagram[*(lime)]{1}_{(-+)} +{\bf{1}}_{(++)}\; \right) \notag \\[5mm]
& \quad = \quad  \ydiagram[*(lime)]{4}_{(++)} + \ydiagram[*(lime)]{3}_{(-+)} + \ydiagram[*(lime)]{2}_{(++)} \notag \\
& \quad \quad + \ydiagram[*(lime)]{1}_{(-+)} \notag \\
& \quad \quad  + {\bf{1}}_{(++)}
\end{align}
(vi)\ {\bf{8}} rep.
\begin{align}
 \ydiagram[ *(lime) ]{1, 1}*[*(white)]{1+1} 
%& \quad = \quad   \ydiagram[ *(lime) ]{1, 1}*[*(white)]{1+1} \otimes {\bf{1}} \notag \\[5mm]
& \xRightarrow{SU(2)}\left( \; \ydiagram[*(lime)]{1}*[*(white)]{1+1}_{(+-)} +\ydiagram[]{1}_{(--)} 
+ \ydiagram[*(lime)]{1}_{(--)} +{\bf{1}}_{(+-)}\; \right) \otimes {\bf{1}}_{(++)}\notag \\[5mm]
&\quad = \quad \ydiagram[*(lime)]{1}*[*(white)]{1+1}_{(+-)} +\ydiagram[]{1}_{(--)} + \ydiagram[*(lime)]{1}_{(--)} +{\bf{1}}_{(+-)}
\end{align}
(vii)\ {\bf{6}} rep.
\begin{align}
 \ydiagram[]{2}
& \xRightarrow{SU(2)}\left( \; \ydiagram[]{2}_{(++)} +\ydiagram[]{1}_{(-+)} +{\bf{1}}_{(++)}\; \right) 
\otimes {\bf{1}}_{(++)}\notag \\[5mm]
&\quad = \quad \ydiagram[]{2}_{(++)} +\ydiagram[]{1}_{(-+)} +{\bf{1}}_{(++)}
\end{align}
(viii)\ $\overline{\bf{6}}$ rep.
\begin{align}
 \ydiagram[*(lime)]{2,2}
& \xRightarrow{SU(2)}\left( \; \ydiagram[*(lime)]{2}_{(++)} +\ydiagram[*(lime)]{1}_{(-+)} +{\bf{1}}_{(++)} \; \right) 
\otimes {\bf{1}}_{(++)}\notag \\[5mm]
&\quad = \quad \ydiagram[*(lime)]{2}_{(++)} +\ydiagram[*(lime)]{1}_{(-+)} +{\bf{1}}_{(++)}
\end{align}
In summary, each $SU(3)$ representations from the 4-rank totally symmetric representation of $Sp(6)$ are decomposed 
into $SU(2)$ representations below. 
\begin{align}
	{\bf{27}} &\xRightarrow{SU(2)} {\bf{5}}_{(++)}\ +\ {\bf{4}}_{(-+)}\times 2\ +\ {\bf{3}}_{(++)}\times 3 \ 
	+\ {\bf{2}}_{(-+)}\times 2\ +\ {\bf{1}}_{(++)}, \notag \\
\bf{24},\overline{\bf{24}}  &\xRightarrow{SU(2)} {\bf{5}}_{(+-)}\ +\ {\bf{4}}_{(--)}\times 2\ +\ {\bf{3}}_{(+-)}\times 2 \ 
+\ {\bf{2}}_{(--)}\times 2\ +\ {\bf{1}}_{(+-)}, \notag \\
%\overline{\bf{15}} &\xRightarrow{SU(2)} {\bf{5}}_{(++)}+{\bf{4}}_{(-+)}+{\bf{3}}_{(++)} +{\bf{2}}_{(-+)}+{\bf{1}}_{(++)} \notag \\
\bf{15},\overline{\bf{15}}  &\xRightarrow{SU(2)} {\bf{5}}_{(++)}\ +\ {\bf{4}}_{(-+)}\ +\ {\bf{3}}_{(++)} \ 
+\ {\bf{2}}_{(-+)}\ +\ {\bf{1}}_{(++)}, \notag \\
\bf{8} &\xRightarrow{SU(2)} {\bf{3}}_{(+-)} \ +\ {\bf{2}}_{(--)}\times 2\ +\ {\bf{1}}_{(+-)} \notag \\
\bf{6},\overline{\bf{6}} &\xRightarrow{SU(2)} {\bf{3}}_{(++)} \ +\ {\bf{2}}_{(-+)}\ +\ {\bf{1}}_{(++)}. 
\end{align}

%%%%%%%%%%%%%%%%%%%%%%%%%%%%%%%%

\end{document}